\documentclass[12pt,english]{article}
\usepackage[T1]{fontenc}
\usepackage[latin9]{inputenc}
\usepackage{amsmath}
\usepackage{amsfonts}
\usepackage{natbib}
\usepackage{amsthm}
\usepackage{a4wide}
\usepackage{setspace}
\usepackage{bbm}
\PassOptionsToPackage{hyphens}{url}
\usepackage{hyperref}
\usepackage{geometry}
\usepackage{enumitem}
\usepackage{comment}
\newtheorem{lem}{Lemma}
\newtheorem{fact}{Fact}
\newtheorem{assume}{Assumption}
\newtheorem{corollary}{Corollary}
\newtheorem{hypothesis}{Hypothesis}
\newtheorem*{hypothesis*}{Hypothesis}


\newtheorem{prop}{Proposition}
\newcommand{\ot}
{\tfrac{1}{2}}
\newcommand{\mcp}{\mathcal{P}}
\newcommand{\kL}{\kappa(0,L,\tilde{q})}
\newcommand{\mcu}{\mathcal{U}}
\newcommand{\mcf}{\mathcal{F}}

\usepackage{pgfplots}

\makeatother

\selectfont
\pgfplotsset{compat=1.18}
\begin{document}

\title{
\textbf{Truth, Lies, and Social Ties:\\ When Image Concerns Fuel Fake News}\thanks{Sisak: Erasmus University Rotterdam and Tinbergen Institute. E-Mail: \href{mailto:sisak@ese.eur.nl}{sisak@ese.eur.nl}. Denter: Universidad Carlos III de Madrid. E-Mail: \href{mailto:pdenter@eco.uc3m.es}{pdenter@eco.uc3m.es}. \textcircled{r} indicates randomized order of author names. We thank Paul Bose, Micael Castanheira, Zuheir Desai, Paul Heidhues, Patricia Himml, Vladimir Karamychev, Jori Korpershoek, David Levine, Agustina Mart\'{i}nez, Kirill Pogorelskiy, Johannes Schneider, Marco Schwarz, Lotte Swank, Otto Swank and the participants of  MAPE  2021,
PET 2022, 
 KVS New Paper Sessions 2022,  the CEPR and Dortmund Political Economy Workshop 2023,
 the 2023 UC3M Micro Retreat, the 2024 Leuphana Microeconomics Workshop,  EPCS  2024, the 2024 ETH/CEPR Workshop on Democracy, SAEe 2024,   Political Economy: Theory Meets Empirics 2025, as well as the seminar audiences at Basel, Rotterdam, and King's College London for helpful suggestions. Philipp Denter gratefully acknowledges financial support by the Agencia Estatal de Investigacion through grants CEX2021-001181-M and PID2022-141823NA-I00 as well as by the  Consejeria de Educacion, Juventud y Deportes de la Comunidad de Madrid through grant CAM - EPUC3M11 (V PRICIT).}}
\author{Dana Sisak \,\textcircled{r}\, Philipp Denter} 

\maketitle

\vspace{-1cm}



\begin{abstract}
 
We study how social image concerns shape information sharing among peers. Individuals receive a signal about a binary state with an observable direction and a veracity that can be learned only through costly verification, which depends on type. We analyze two motives: a desire to appear competent and a desire to signal one's worldview. For each, we characterize equilibrium sharing, derive implications for the relative diffusion of true versus false news, and compare welfare. The two motives can both generate greater sharing of false than factual news but imply distinct, testable patterns which we relate to existing empirical evidence.

\end{abstract}



\smallskip

 \newpage

\section{Introduction}
Billions of people around the world share information with their friends, families, and other peers through social media platforms. The rise of these platforms has significantly reduced the cost of information sharing, making it easier for individuals to connect and interact with others.\footnote{Research has documented the importance of information sharing in various contexts, including the adoption of microfinance (e.g., \citealp{BanerjeeEtAl:2013}), vaccination campaigns (e.g., \citealp{BanerjeeEtAl:2019}), and electoral behavior (e.g., \citealp{allcott2017social}; \citealp{pogorelskiy2019news}).}
However, concerns about the quality of shared information and its societal consequences have been mounting in recent years. This is due to the increasingly widespread phenomenon of disinformation, which has been shown to undermine trust in democratic institutions\footnote{See, for example, \url{https://www.brookings.edu/blog/fixgov/2022/07/26/misinformation-is-eroding-the-publics-confidence-in-democracy}.} and to influence electoral outcomes (\citealp{zimmermann2020mistrust}; \citealp{cantarella2023does}).\footnote{See also \url{https://www.theguardian.com/world/2019/oct/30/whatsapp-fake-news-brazil-election-favoured-jair-bolsonaro-analysis-suggests}.} Disinformation has also contributed to public mistrust in vaccines during the COVID-19 pandemic, leading to suboptimal vaccine uptake in many countries (\citealp{MontagniEtAl:2021}).

Unfortunately, the problem of disinformation is not confined to those who create it. Peer-to-peer sharing can dramatically amplify the reach and impact of both factual and false information. Recent research shows that disinformation spreads ``\textit{farther, faster, deeper, and more broadly}'' than factual content on platforms like Twitter/$\mathbb{X}$ \citep{VosoughiRoyAral:2018}, suggesting that the diffusion of fake news is driven to a great extent by everyday users.

To develop effective remedies against the spread of disinformation and its consequences, it is essential to understand the underlying factors that lead individuals to share information.
One important reason why individuals share information is to influence their social image, as shown by \cite{LeeMaGoh:2011}, \cite{LeeMa:2012}, and \cite{KuempelEtAl:2015}. By sharing exclusively high-quality information, individuals can signal that they are \emph{able}---that is, capable of discerning valuable information from dubious content. Conversely, sharing information that later turns out to be false can damage that image and lead to a loss of status. However, information sharing is not only a signal of ability; it can also serve to communicate one's \emph{worldview}. Sharing content with a particular slant or direction can signal ideological alignment and indicate support for the perspective it promotes.

In this paper, we build a theoretical model to analyze the conditions under which individuals share information and the resulting consequences when they have social image concerns. We assume an unknown binary state of the world, $\omega$, which could represent, for example, whether human activity drives climate change or whether a vaccine is effective. A sender ($S$) receives a binary signal $\sigma$ about $\omega$ and can choose whether to share it with a receiver ($R$). The signal has two dimensions. First, it contains a headline, for instance, one that supports or opposes the idea of human-induced climate change. This headline may be surprising---contradicting prior beliefs---or unsurprising. Only surprising signals, provided they are sufficiently informative, influence the receiver's decision-making. We refer to this first dimension as the signal's \emph{relevance}, which is immediately observable to both sender and receiver at no cost.

Second, the signal may be either proper, meaning it is fact-based, or fake, in which case it is fabricated and therefore false. We call this dimension \emph{veracity}. Proper signals are correlated with the true state of the world and thus informative, whereas fake signals convey no information. Determining a signal's veracity is possible but costly, and only individuals with high ability find it worthwhile to do so.

The sender is motivated by social image concerns: she cares about how she is perceived by her peers. We examine the two types of social image concerns mentioned earlier. First, the desire to be seen as \textit{able}, which can be demonstrated by accurately distinguishing between proper and fake signals. Second, the desire to be perceived as holding a particular \textit{worldview}.

The model is designed to capture a common scenario: an individual deciding whether to share, for example, a newspaper article with another person or a broader audience on social media. The article's relevance can typically be inferred from its headline alone. Verifying its veracity, however, requires reading the article and engaging with its content---an effort that demands time, attention, and the ability to evaluate arguments critically.

We characterize equilibrium information-sharing behavior driven by each of the two social image motives. Under the ability motive, high-ability senders filter out disinformation and share only signals that are both relevant and based on facts. In contrast, low-ability senders cannot reliably identify fake signals and therefore randomize between sharing relevant content and choosing not to share at all. When the motive is to signal one's worldview, the sharing pattern changes: only individuals with strong or extreme beliefs are willing to share corresponding signals, while moderates refrain from sharing any signal.

Our main results concern the conditions under which the quality of information deteriorates through the sharing process.

    \begin{itemize}
        \item 
        With an \emph{ability} motive, fake news may be shared disproportionately when improper signals are predominantly \emph{surprising} to receivers. They are shared especially by \emph{low-ability} senders who hold \emph{different worldviews} than their audience. In such cases, senders tend to share only  signals that are surprising to receivers and withhold unsurprising ones. Lower sharing costs tend to exacerbate this pattern, making the disproportionate spread of fake news more likely. This may especially hurt low-ability receivers. At the same time, the quality of news \emph{conditional on relevance} always improves under this motive. 
        
        \item With a \emph{worldview} motive, fake news may be shared disproportionately when improper signals predominantly \emph{align} with the receivers' prior beliefs. They are shared especially by senders who share \emph{a similar worldview} as their audience. In such cases, senders are more likely to share signals that confirm the receivers' expectations and withhold those that contradict them. Lower sharing costs encourage sharing by senders with moderate worldview with no clear effect on fake news propagation. Since, conditional on relevance there is no filtering of signals, sharing costs tend to decrease decision quality of both types of receivers by restricting access to information.
        
    \end{itemize}

A deterioration of information quality after sharing may arise regardless of the underlying social image motive and reveals a fundamental vulnerability in peer-to-peer information transmission. At the same time, when fake news is shared disproportionately, the two social image motives give rise to clearly distinct sharing patterns. As a result, our model generates novel, testable predictions that allow researchers to empirically distinguish between different types of social image concerns, which is important to find effective remedies against the spread of disinformation. The patterns we uncover theoretically align with empirical findings (see Section~\ref{sec:discussion}).

Our findings contribute to the ongoing debate on how to reduce fake news sharing on social media. \cite{PennycookEtAl:2021} and \cite{GurievHenryMarquisZhuravskaya:2023} show that nudging users to focus on accuracy is an effective way to lower the spread of disinformation while increasing the spread of proper information.\footnote{In fact, \cite{GurievHenryMarquisZhuravskaya:2023} show that all of their policy interventions (additional click, fact check available, first assess veracity, accuracy prime) predominantly work though increasing the salience of reputation.} Our model shows that this may not always be effective, at least in increasing the overall quality of shared news. While in  contexts where fake news is shared disproportionately due to a worldview motive,  nudging towards accuracy---and thus an ability motive---may work, in other contexts prompting ability signaling decreases the quality of information.\footnote{Moreover, the supply and characteristics of fake news are not fixed and may adapt to the prevailing social image motive.} In addition, our model cautions that to keep the accuracy motive salient, senders need to believe that receivers will consistently engage with the information shared. 

\paragraph{Literature:}
This paper contributes to several literatures both in political economy and microeconomic theory. 
One related strand studies peer-to-peer sharing of non-falsifiable information. 
A closely related paper from this literature is due to \cite{CheKartik:2009}, who analyze a setting in which a sender decides whether to disclose a piece of information about a binary state to a receiver. 
In their model, the purpose of information sharing is not to gain social image or status utility---which generates incentives to share information in the present paper---but rather to influence the receiver's action. 
\cite{pogorelskiy2019news} show that voters are more likely to share political news with their peers if the news  aligns with their preferences.
\cite{LEVY2018262} and \cite{BowenEtAl:2023} examine the consequences of peer-to-peer information sharing for belief polarization in society. 
\cite{LEVY2018262} show how information sharing can lead to polarization when individuals update according to what they call the ``Bayesian Peer Influence heuristic,'' while \cite{BowenEtAl:2023} demonstrate that even minor misperceptions about the signal-generating process can render collective learning inefficient and result in polarization of beliefs. 
While similar in spirit, in these papers information sharing is non-strategic. 
By contrast, our results are driven by individuals' desire for social image or status utility, which shapes their information-sharing behavior.

This paper also fits into the growing theoretical literature on the propagation of fake news. \cite{GrossmanHelpman:2019} study a model of electoral competition in which political parties can spread fake news about policy positions. They abstract from the sharing of such information among peers as a main driver of the spread of fake news, and sharing serves primarily to exert influence.
\cite{AcemogluEtAl:2010}, \cite{Papanastasiou:2020}, \cite{DDG:2021}, \cite{DenterGinzburg:2026}, and \cite{germano2026ranking} study how misinformation and fake news spread through social networks, though they do not explicitly model the sharing decision.

In contrast, \cite{KrantonMcAdams:2024} assume that an individual shares information if and only if she believes, with sufficiently high probability, that the information is genuine.
Similarly, \cite{AcemogluOzdaglarSiderius:2023} study the decision to dislike or share potentially fake information with one's peers. Disliking a piece of information brings utility if it turns out to be fake, whereas sharing brings utility if the information turns out to be factual or if peers decide to reshare it. If peers decide to dislike a shared piece of information, one incurs disutility.
Unlike in \cite{KrantonMcAdams:2024} and \cite{AcemogluOzdaglarSiderius:2023}, in our paper agents are both horizontally (by ideology) and vertically (by ability) differentiated, and we consider two distinct social-image motives. Moreover, our analysis reveals that the veracity of a piece of information may be orthogonal to it being shared.


One focus of our paper is to determine the circumstances that lead to the sharing of fake news. There exists a wide empirical literature concerned with this questions.
\cite{GuessEtAl:2019} show
that over 65 year old individuals  and conservative-leaning individuals were most likely to share (mostly pro-Trump) fake news. 
\cite{PennycookEtAl:2021} argue that various motives matter for sharing information on Twitter.
When signaling one's partisanship is  salient, copartisan fake news are spread knowingly,
but if accuracy becomes more important, fake news sharing is reduced.\footnote{This is in line with recent evidence that people are reasonably good at detecting real from fake news \citep{AngelucciPratt:2021}.} 
Similarly, \cite{OsmundsenEtAl:2021} find that fake news sharing is often driven by partisan sentiment, rather than by individuals' ignorance regarding the veracity of a news item.
\cite{GurievHenryMarquisZhuravskaya:2023} 
show  that ability signaling, a desire to persuade receivers, and signaling of partisanship are all  drivers of information sharing. Moreover, they show  that priming the circulation of fake news is an effective tool to reduce sharing  of fake news while preserving the  dissemination of genuine news.
Just as \cite{PennycookEtAl:2021} and \cite{GurievHenryMarquisZhuravskaya:2023}, we  study the motives of ability  and partisanship signaling, using a formal model. Unlike these papers, we model active receivers who may engage with the shared news to ascertain its quality (``fact checking'') and then form beliefs about the type of the sender in Bayesian fashion. Interestingly, we find that even with an accuracy motive, fake news may be spread disproportionately, as found by \cite{VosoughiRoyAral:2018}. We identify and characterize settings in which fake news sharing is most problematic with each motive and derive empirical predictions that can help to distinguish between the different motives.

\section{Model}
\label{sec:model}

\paragraph{Players.}
There are two roles in the game: a sender, $S$, and one or more receivers, $R$. A sender $S$ receives a signal $\sigma$ about an unknown state of the world and decides whether to fact-check and share it. We use $R$ to denote a generic receiver. Depending on the application, the sender communicates either with a single receiver, as in a private message on Twitter/$\mathbb{X}$, WhatsApp, or Telegram, or with a large audience, as in a public post or group message on these platforms. We approximate the latter by a continuum of receivers of mass one.

\paragraph{State of the World.}
At the beginning of the game, nature draws the state of the world
$\omega\in\{0,1\}$. The true probability that $\omega=1$ is
$\mathbb{P}[\omega=1]=p_T\in(0,1)$, while $\omega=0$ occurs with
complementary probability $1-p_T$. Players need not share this objective
probability and may hold heterogeneous prior beliefs about the state.

\paragraph{Signal Generating Process.}
Nature then generates a signal $\sigma\in\{0,1\}$ that is received by the sender. Each signal has a veracity status $\nu\in\{0,1\}$, where $\nu=1$ denotes a \emph{proper} signal and $\nu=0$ a \emph{fake} signal. With probability $1-q$, the signal is proper, $\nu=1$, meaning that it is based on facts and informative about the state. With probability $q\in(0,1)$, the signal is fake, $\nu=0$, and carries no information about $\omega$.

If $\nu=1$, the signal matches the state with probability
\[
\mathbb{P}[\sigma=1\mid\omega=1,\nu=1]
=
\mathbb{P}[\sigma=0\mid\omega=0,\nu=1]
=
\eta\in\left(\ot,1\right],
\]
and is incorrect with complementary probability $1-\eta$. We refer to $\eta$ as the \emph{precision} of a proper signal.
If $\nu=0$, the signal is independent of the state. In particular,
\[
\mathbb{P}[\sigma=1\mid\omega=1,\nu=0]
=
\mathbb{P}[\sigma=1\mid\omega=0,\nu=0]
=
\beta\in[0,1],
\]
while $\sigma=0$ occurs with probability $1-\beta$. We interpret $\beta$ as the \emph{bias} of a fake signal. Such a signal may, for example, have been created by a biased agent attempting to influence players' beliefs. The signal-generating process is common knowledge.

\paragraph{Player Types.}
Before any player observes the signal, her private type $t_i=(p_i,\theta_i)$ is realized. The two components of type are drawn independently, and types are independent across players. Their distributions are common knowledge, while their realizations are private information.

Players differ along two dimensions. First, they may hold different prior beliefs about the state of the world. We assume that an underlying ideological or partisan dimension shapes these beliefs. For example, a player who is generally skeptical of climate change may approach a new environmental policy with skepticism and therefore attach a relatively low prior probability to a favorable state, while a more environmentally minded player may hold the opposite view.\footnote{A behavioral mechanism leading to such divergent priors based on an underlying ideology could be motivated reasoning as, for example, in \cite{TaberLodge:2006}.} For simplicity, we identify this ideological or partisan dimension directly with player $i$'s prior $p_i\in[0,1]$ and refer to it as her \emph{worldview}.\footnote{This assumption helps us save on notation. Introducing instead a separate parameter capturing a player's worldview---for example a fixed ideology as in \cite{buisseret2022polarization}, a bliss point as in the electoral competition literature \citep{downs1957economic}, or a bias as in the cheap-talk literature \citep{CrawfordSobel:1982}---and allowing it to be imperfectly correlated with $p_i$ would not change our results.} Player $i$'s worldview is drawn from distribution $F_i$ with strictly positive density $f_i$ on $[0,1]$, where $i\in\{S,R\}$. Its realization $p_i$ is private information, while $F_i$ is common knowledge.

Second, players differ in their ability to verify information, $\theta_i\in\{L,H\}$. With probability $\lambda_i\in(0,1)$, player $i$ has high ability, $\theta_i=H$, and can verify a signal at zero cost, $c_F(H)=0$. With complementary probability $1-\lambda_i$, she has low ability, $\theta_i=L$, and the cost $c_F(L)$ is prohibitively high. We assume that $\lambda_S+\lambda_R<1$. Ability is private information, while $\lambda_i$ is common knowledge. Thus,
player $i$'s private type is 
$t_i=(p_i,\theta_i)\in\mathcal T_i\equiv[0,1]\times\{L,H\}$.

\paragraph{Fact-checking.}
After observing the signal and before taking any further action, player $i$ may
\emph{fact-check} it. Denote this decision by
$v_i(\sigma,t_i)\in\{0,1\}$, where $v_i=1$ means that player $i\in\{S,R\}$
checks the signal's veracity. Low-ability players cannot fact-check and therefore
choose $v_i=0$ for any signal. High-ability players, in contrast, always
fact-check a signal that is surprising given their worldview $p_i$. We call a signal \emph{surprising} if its realization
runs counter to player $i$'s worldview $p_i$.

Fact-checking reveals the signal's veracity status $\nu$. Denote player $i$'s
information about veracity by $\nu_i\in\{0,1,U\}$, where $U$ indicates that
veracity is unknown. Thus, $\nu_i=\nu$ if player $i$ fact-checks and
$\nu_i=U$ otherwise. 
Finally, let $\tilde q^i(\sigma,p_i,\nu_i)$ denote player $i$'s posterior belief
that the signal is fake. If $\nu_i=U$, the signal realization may nevertheless
be informative, yielding $\tilde q^i(\sigma,p_i,U)
=
\mathbb{P}[\nu=0\mid\sigma,p_i]\in[0,1]$. 
If instead $\nu_i\in\{0,1\}$, veracity is known and hence
$\tilde q^i(\sigma,p_i,\nu_i)=1-\nu_i\in\{0,1\}$.

\paragraph{Sharing and Receiver Actions.}
After observing the signal and any information obtained through fact-checking, the sender decides whether to share it. 
Let $\kappa(\sigma,t_S,\tilde q^S)\in[0,1]$ denote the probability that
sender $S$ shares signal $\sigma$.
Thus, sharing may depend on the signal realization, the sender's type, and any information she obtained about the signal's veracity.

If the signal is shared, receiver $R$ observes its realization $\sigma$, but not the sender's fact-checking decision or the information obtained through fact-checking. She fact-checks according to the rule described above and then chooses an action $a\in\{0,1\}$. The receiver wants her action to match the state of the world and obtains $u^R(a)=-|\omega-a|-v_Rc_F(\theta_R)$. Let $\mathcal I_R$ denote the information available to receiver $R$ after observing a shared signal: the realization $\sigma$, receiver $R$'s own type $t_R$, and any information $\nu_R$ obtained through fact-checking. Then, $R$ chooses $a=1$ whenever $\mathbb{P}[\omega=1\mid\mathcal I_R]\geq \frac12$, and chooses $a=0$ otherwise. We assume that she chooses $a=1$ when indifferent.

\paragraph{Social Image.}
Sharing a signal entails a direct cost $c_S\geq0$. The sender derives no direct utility from informing receivers about the state. Instead, the benefit from sharing comes from \emph{social image}: by deciding which signals to share, the sender can influence receivers' beliefs about her type.

We consider two motives for social image. First, a sender may wish to be perceived as having high ability. In this case, her social image vis-\`a-vis
 receiver $R$ is $R$'s posterior belief that $\theta_S=H$. Second, a sender may care about how accurately receiver $R$ perceives her worldview. Denote by $\widehat p_S=\mathbb{E}[p_S\mid\mathcal I_R]$ receiver $R$'s point estimate of the sender's worldview. A sender with a worldview motive obtains social image utility $-|p_S-\widehat p_S|$ and therefore benefits when the receiver's point estimate is close to her true ideological or partisan type.

With a single receiver, the sender cares about expected social image, whereas with a large audience, she cares about aggregate social image across receivers. Since receiver types in the audience are distributed as a representative receiver and the audience has mass one, aggregate social image coincides with expected social image.

\paragraph{Equilibrium.}
Our focus is on Perfect Bayesian Equilibria of the information-sharing game. The sender first decides whether to fact-check the signal and then whether to share it, maximizing her social-image utility net of fact-checking and sharing costs. In doing so, she anticipates how each receiver type reacts to the shared information and what these reactions imply for beliefs about her type. Receivers, in turn, decide whether to fact-check and choose an action $a\in\{0,1\}$ given the sender's strategy and all information available to them. Beliefs are updated according to Bayes' rule whenever possible and are consistent with equilibrium strategies.

We first analyze the case in which the sender uses information sharing to signal her ability. Section \ref{sec:worldview} then considers a sender whose sharing incentives are instead driven by her worldview.

\section{Information Sharing to Signal Ability}
\label{sec:ability}

\subsection{Equilibrium}

In this section we study situations where a sender wants to signal her ability to recognize improper signals. The sender thus chooses her sharing strategy to maximize the probability that she is perceived as a high ability type by $R$. To focus on the ability motive, we assume that sender types do not also differ in their beliefs $p_S$.\footnote{Alternatively, we can assume that beliefs are common knowledge.} Since worldviews are fixed throughout this section, we suppress them from the strategy notation. For simplicity, we assume that also receiver types do not differ in their beliefs, though we allow for $p_S\neq p_R$. To focus on situations where social image utility \emph{can} be gained, we restrict attention to $\eta\geq \max \{p_R,1-p_R\}$, and thus a proper signal is sufficiently informative to influence the optimal decision of $R$.  These signals can thus be used by $S$ to gain social image utility. Without loss of generality, we assume that $p_R>\ot$, in which case only signal realizations $\sigma=0$ are surprising and thus relevant for the decision of the receivers.\footnote{See Supplementary Appendix A.1 for a formal proof of this statement as well as a result about equilibrium sharing behavior if this condition is violated. The intuition is straightforward. If the signal is not surprising, or if the quality of a proper signal is too low, $\eta<p_R$, then independent of the signal's veracity, the optimal action of the receiver is $a=1$. However, when the signal is surprising and $\eta\geq p_R$, it has the potential to change a receiver's optimal action. If improper though, $R$ better disregard it and take the decision based solely on the prior.}

As typical, there are multiple Perfect Bayesian Equilibria of the game. However, in our framework standard equilibrium refinements such as the Intuitive Criterion or D1 cannot help to  narrow down the set of possible equilibria. Our interest here is in equilibria where social image utility is gained through signaling of ability. This can be done both by identifying and sharing proper signals, and by identifying and sharing fake signals.\footnote{In Section A.1 of the Supplementary Appendix we discuss which other types of equilibria cannot exist.} We assume social image utility is gained through the sharing of proper signals. Then, a high ability sender has a strictly greater incentive to relay a proper and surprising signal than a fake and surprising signal, because the latter will be interpreted by high ability receivers as evidence that the sender has  low ability. Therefore, we should expect the high ability sender to choose $\kappa(0,H,1)=0$ and never relay surprising, improper signals. The signal that should yield the greatest expected status gain is a surprising and proper signal, and hence we should expect that these signals are always shared, $\kappa(0,H,0)=1$. 

At the same time, it is unclear whether $S$ should withhold or share a signal that is not surprising. Indeed, there are generally multiple equilibria of the game where social image utility can be gained from sharing proper signals, the major difference between them being if unsurprising signals are shared. In the following, we  focus on the equilibrium in which $S$ does not share \emph{any} non-surprising signals.\footnote{This equilibrium is the ex-ante preferred one from the sender's perspective when $c_S>0$. Bayes-plausibility implies that, in expectation, status remains unchanged. This means that the more signals are shared, the lower the expected utility of the players because expected sharing costs increase. Moreover, equilibria where non-surprising signals are shared with positive probability are less likely to feature a deterioration in the quality of information. Hence, our analysis identifies a worst-case scenario in line with our objective of identifying situations where fake news sharing is most problematic.}

\begin{prop}[Ability Signaling]
\label{prop:EQ}
There exist thresholds
$0 < c_L < c_H < 1$, a strictly decreasing function
$\bar{q}: [0,c_L] \to [0,1]$ with $\bar{q}(0) = 1$ and $\bar{q}(c_L) = 0$,
and an off-equilibrium belief threshold $\bar{\pi}_1\in(0,1)$ such that the following
Perfect Bayesian Equilibrium exists for any
$\pi_1\leq\bar{\pi}_1$:
\begin{itemize}
  \item No non-surprising signals are shared. Formally,
   $\kappa^*(1,H,\tilde{q}^S) = \kappa^*(1,L,\tilde{q}^S) = 0$.
\item If $c_S\leq c_H$, a high ability sender checks every $\sigma=0$ signal,
shares every proper one, and never shares a fake one. Formally, $v_S^*(0,H) = 1$, $\kappa^*(0,H,0) = 1$,  and
  $\kappa^*(0,H,1) = 0$. If $c_S>c_H$, a high ability sender  shares no signal.
  \item If $c_S < c_L$ and $q \le \bar{q}(c_S)$, a low ability
  sender shares a surprising signal with probability
  $\kappa^*(0,L,\tilde{q}^S(0,p_S,U)) \in (0,1)$. If $c_S\geq c_L$ or $q > \bar{q}(c_S)$, a low ability sender shares no information.
\end{itemize}
\end{prop}

When the share of improper signals $q$ and sharing costs $c_S$ are not too large, sharing a surprising signal with
positive probability is beneficial for a low ability sender in equilibrium.
Moreover, as long as $c_S < c_L$, the probability that the low ability type
shares a surprising signal, $\kappa_0^*$, decreases in $c_S$. Intuitively, a
higher $c_S$ makes sharing less attractive relative to remaining silent. A
lower $\kappa_0$ restores the balance: it raises the status from sharing while
lowering the status from silence, leaving the low ability sender indifferent
between the two.
Moreover, $\kappa_0^*$ also decreases in $p_S$. What is the reason for this
finding?

\begin{fact}
The belief of a sender that a given signal $\sigma$ is fake increases in
$|p_S-\sigma|$.
\label{fact:1}
\end{fact}

As $p_S$ increases, the low ability sender's belief that her signal
$\sigma=0$ ($\sigma=1$) is fake increases (decreases), which makes sharing
less (more) attractive.\footnote{The proofs of both comparative statics
results are contained in the proof of Fact \ref{fact:1}.} All other
comparative statics are not as clear cut.

The low ability sender's suspicion of her own signal also explains why she is
the first to fall silent as sharing becomes more expensive. Once $c_S$ reaches
$c_L$, she stops sharing altogether while the high ability sender continues.
The reason is that she cannot tell what she is passing on: some of the signals
she relays are fake, and she loses social image whenever a receiver verifies
one. Having checked before sharing, the high ability sender never bears this
risk. Sharing is therefore worth less to the low type at any given cost.

\subsection{The Quality of Shared Information}

We now study the quality of shared information. 
We measure quality as the fraction of shared information that is fake. Defining the probability of proper news being shared as $\sigma^\mathcal{P}$ and the probability of fake news being shared as $\sigma^\mathcal{F}$, implies that our measure of the expected quality of shared information equals
\begin{equation}
\label{eq_defin_gamma}
\gamma\equiv \frac{\sigma^\mathcal{F}}{\sigma^\mathcal{F}+\sigma^\mathcal{P}}.
\end{equation}
Naturally, when $\gamma$ decreases, we interpret this as increasing quality, because the probability that a shared signal is fake is lower. Similarly,  larger $\gamma$ means the quality of shared information decreases. 


We first study how the availability of social media platforms, which facilitate easy information sharing and therefore decrease the cost of sharing information with our peers, affects the quality of shared information. Many researchers have attributed increasing spread of misinformation to exactly those platforms. The next result shows that decreasing sharing cost indeed increases the share of disinformation 
of shared content. It also speaks to the effect of worldview polarization, as measured by the worldview of the sender $p_S$ relative to the receiver(s) $p_R$ (which by assumption $p_R>\frac{1}{2}$):

\begin{prop}
\label{prop:dgammadc}
Consider the equilibrium described in Proposition \ref{prop:EQ}. $\gamma$ strictly decreases in both $p_S$ and $c_S$ for $c_S\in[0,c_L)$ and equals zero for $c_S\in[c_L,c_H]$.\end{prop}

Sharing costs affect the spread of fake information through low-ability senders, since high-ability senders never share fake information. A low-ability sender randomizes between sharing surprising signals and keeping them to herself. When sharing becomes cheaper, it becomes, ceteris paribus, more attractive to share. Increasing $\kappa_0$ reduces the status gain from sharing and raises the status gain from not sharing, restoring the balance between relaying and withholding information. As a result, higher sharing costs reduce $\gamma$ and improve the quality of shared information.

Also the sender's worldview $p_S$ matters for the quality of information shared. The greater the sender's belief that  $\omega=1$, and thus the higher $p_S$, the lower $\kappa_0^*$, as explained above. Thus, low types are more conservative in their sharing decision, which improves the quality of information shared. Consequently, $\gamma$ decreases in $p_S$. Recall that $p_R>\ot$. Then, $\gamma$ is highest when $p_S \rightarrow 0$, and thus the sender and the receiver(s) maximally disagree in their belief about the state $\omega$.

\begin{figure}
  \centering
  \includegraphics[width=\textwidth]
  {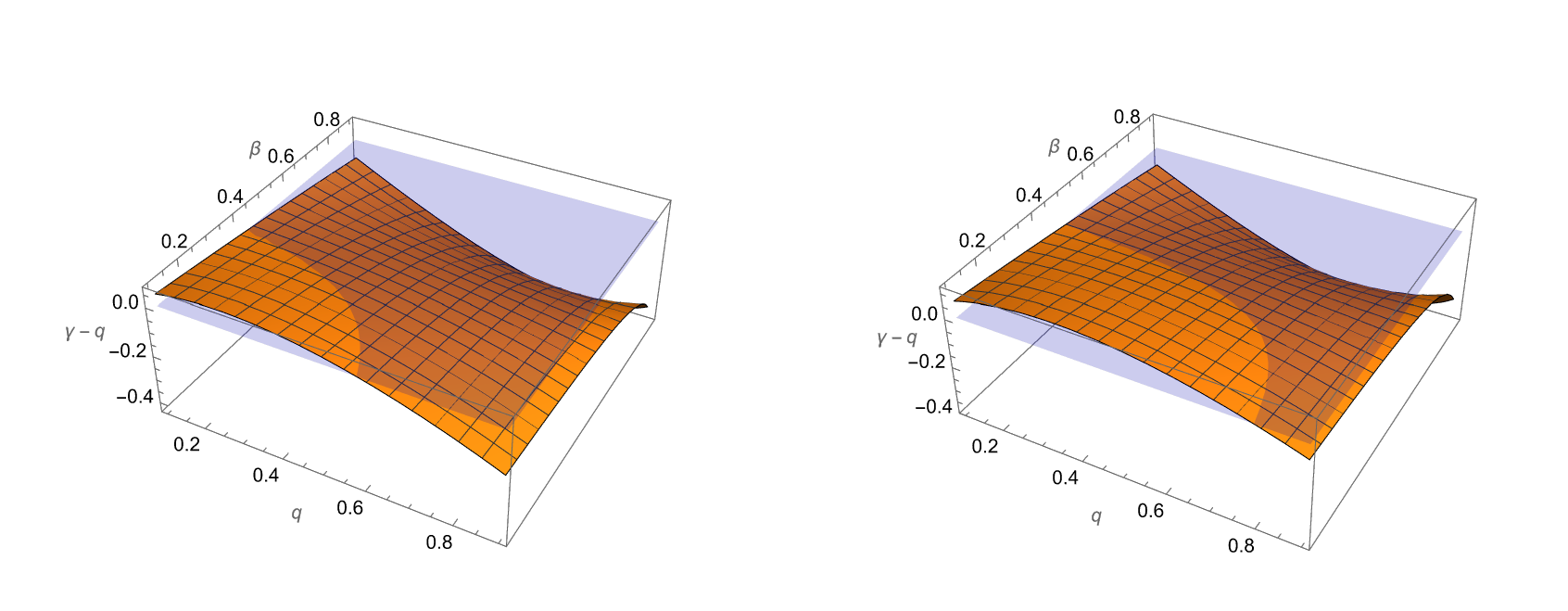}
  \caption{$\gamma-q$ as a function of $q$ and $\beta$ when $\lambda_S=\frac{1}{5}$ (left panel) and $\lambda_S=\frac{1}{10}$ (right panel), as well as $c_S=0$, $\eta=\frac{2}{3}$, $\lambda_R=\frac{1}{5}$, and $p_S=p_R=p_T=\frac{2}{3}$. The light-shaded blue plane divides the positive and negative halfspace and marks $\gamma-q=0$. }\label{fig_share}
\end{figure}

But how low can the quality of information after sharing become? Is the fraction of shared information that is fake always smaller than the expected fraction of fake information received by the sender, $\gamma<q$? This would imply that the quality of shared information increases due to filtering by senders. This is not necessarily the case. 
Figure \ref{fig_share} shows how $\gamma-q$ changes as a function of $q\in[\frac{1}{10},\frac{9}{10}]$,  $\beta\in[\frac{1}{10},\frac{9}{10}]$, and $\lambda_S\in\{\frac{1}{10},\frac{1}{5}\}$ when $c_S=0$, $\eta=\frac{2}{3}$, $\lambda_R=\frac{1}{5}$, and $p_S=p_R=p_T=\frac{2}{3}$.\footnote{For these parameters, $0<\kappa_0^*<1$ and thus the equilibrium characterized in Proposition \ref{prop:EQ} exists.} We can see that in the example information quality deteriorates after sharing, $\gamma>q$, when $q$ and $\beta$ are small and thus fake news are not prevalent and biased towards the surprising state. Moreover, when the share of high ability senders $\lambda_S$ decreases, the parameter range such that $\gamma>q$ increases. To shed further light on which situations are problematic, the next proposition offers a sufficient condition for the quality of information to improve after sharing.

\begin{prop}
\label{prop:dgammadbeta}
Consider the equilibrium identified in Proposition \ref{prop:EQ}. 
A sufficient condition for $\gamma<q$ is $c_S<c_L$ and
\[
\beta>1-\frac{p_T(1-\eta)+(1-p_T)\eta}{1-\lambda_S},
\]
or $c_S\in[c_L,c_H]$.
\end{prop}

Thus, only when fake news are biased towards the surprising state, i.e., if $\beta$ sufficiently small, is there scope for information quality deterioration. The same holds for sufficiently many low-ability senders, that is, $\lambda_S$ small, because its them who share fake signals, as well as a high true probability that the state is 1 ($p_T$ large) and thus surprising signals $\sigma=0$ are likely fake.\footnote{For example, when $\lambda_S=\frac{1}{10}$, $\eta=\frac{2}{3}$ and $p_T=\frac{2}{3}$, we get a sufficient condition of $\beta>\frac{41}{81}=0.506$ (right graph) while when $\lambda_S$ increases to $\frac{1}{5}$, we get $\beta>\frac{4}{9}=0.444$ (left graph).} 
To conclude, the quality of information shared relative to information received tends to be worse when $\beta$, $p_S$, $\lambda_S$ and $c_S$ are low.


While we established that the quality of information after sharing can deteriorate under an ability motive,
our model also highlights the limitations of this measure. Receivers in our setting only care about relevant signals, as only those are pivotal for their decision. The quality of relevant signals unambiguously increases under the ability motive, due to filtering by high-ability senders. The decrease in quality is thus an artifact of the filtering of signals by \emph{relevance}. This is in line with the empirical evidence in \cite{VosoughiRoyAral:2018} who show that fake news, especially in the political domain, spread ``\textit{farther, faster, deeper, and more broadly}'' on Twitter/$\mathbb{X}$ than factual news.\footnote{While they consider the differential spread of false relative to true news, we consider the quality of information before and after sharing. If fake signals are reshared more often than proper ones, this is equivalent to the quality of information deteriorating after sharing.} They also find that false news are more novel than proper news---$\beta$ is small relative to $p_T$ in terms of our model---and that people were more likely to share novel information, which is in line with the equilibrium in Proposition \ref{prop:EQ}.

\subsection{Welfare}
We now examine how higher sharing costs affect receiver welfare. By discouraging low-ability senders from sharing, these costs raise information quality. Also empirically such a policy has been shown to be effective in reducing the sharing of fake news \citep{GurievHenryMarquisZhuravskaya:2023}.\footnote{We provide a formal analysis to complement this discussion as well as additional results on sender welfare in our Supplementary Appendix A.2.} Since fact-checking is costless  for high ability $R$, our measure of receiver welfare is the probability of a correct choice. Because high ability receivers fact check relevant signals they are  hurt by an increase in sharing costs, because such a policy also weakly decreases the amount of proper signals shared with them. The absolute number of proper and surprising signals shared with them is the sole determinant of their welfare. 

To the contrary, for low ability receivers, an increase in sharing costs may be \textit{welfare improving}. A necessary condition for this is that $c_S<c_L$ and that $R$ finds surprising signals shared with her sufficiently informative, leading her to change her optimal decision. When signals shared by low ability $S$ are of low quality, because fake news are relatively prevalent ($q$ is high) and/or biased to be surprising to $R$ ($\beta$ is small), higher sharing costs may improve low ability $R$'s decisions in expectation. Thus, a policy affecting sharing costs is most likely welfare improving if there are many low ability $R$ and $S$, shared signals by low ability $S$ are of bad quality but low ability $R$ still sufficiently trust the information they obtain to use it in their decision making.

\section{Information Sharing to Signal Worldview}
\label{sec:worldview}
\subsection{Equilibrium}
To study information sharing as a means of signaling worldview, we now relax the assumption that the sender's worldview $p_S$ is common knowledge. 
Instead,  the sender's prior belief is distributed according to $F_S$, as introduced in Section~\ref{sec:model}. 
We interpret $S$'s desire to signal her worldview to $R$ as the wish to induce in $R$ an estimated worldview $\widehat{p}_S$ that is as close as possible to her actual worldview. 
Her social image utility equals
\[
u^S(p_S)
= -\int_{0}^{1} \left| p_S - \widehat{p}_S(p_R) \right| \, dF_R(p_R),
\]
where $\widehat{p}_S(p_R)$ denotes the perceived worldview of $S$ by a receiver with worldview $p_R$.

As in most games of information transmission, there exists an equilibrium in which receivers ignore any information sent by $S$ and do not update their beliefs about the sender's worldview. Hence, in turn, it is optimal for $S$ not to send any signal. We instead focus on equilibria in which $R$ responds to the sender's sharing decision. We refer to such an equilibrium as a \emph{Responsive Equilibrium}. We focus on equilibria with natural language: sharing $\sigma=1$ should weakly increase $R$'s estimate of the worldview of $S$, while sharing $\sigma=0$ should weakly decrease it, relative to not sharing anything.

While high-ability senders may learn the veracity of surprising signals, veracity does not confer differential social image utility, as we argue below. Hence, although fact-checking may occur, $S$'s sharing decision does not condition on veracity.\footnote{To illustrate why it is not possible to gain differential social image utility from proper versus improper signals, assume that all senders and receivers observe veracity at no cost. Equilibrium then requires two versions of $C_l$ as defined below, one for proper signals with threshold $p^p_{Sl}$ and one for improper ones with threshold $p^i_{Sl}$, and analogously for $C_h$. Since social image utility from not sharing, $\widehat p_S(\emptyset)$, is identical across all four conditions, and as there are no differential costs of sharing proper versus improper signals, it follows that $p^p_{Sl}=p^i_{Sl}$ and $p^p_{Sh}=p^i_{Sh}$. Note that this would be different in a mixed-motive setting as discussed in Section \ref{sec:mixed_motives}, where both ability and worldview social image utility can be gained. As improper signals are costly on the ability dimension, they could act as stronger signals of worldview.}
Therefore, for forming beliefs about $S$'s worldview, $R$ notes only whether a signal is sent and what the ``headline'' of the signal is, $\sigma\in\{0,1\}$.



We will focus on responsive equilibria of the following form:
\begin{itemize}
    \item A sender that receives a signal $\sigma=0$ shares the signal if and only if her worldview $p_S$ lies in $[0,p_{Sl}]$ for some $p_{Sl}\in(0,1)$;
    \item A sender that receives a signal $\sigma=1$ shares the signal if and only if her worldview $p_S$ lies in $[p_{Sh},1]$ for some $p_{Sh}\in(0,1)$.
\end{itemize}

Denote by $
\widehat{p}_R = (1-q)\big(\eta p_R + (1-\eta)(1-p_R)\big) + q\beta$
a receiver's belief about the probability that the sender receives a signal $\sigma=1$. 
In a \emph{Responsive Equilibrium}, $R$'s posterior beliefs about the worldview of the sender---after observing either $\sigma=0$, $\sigma=1$, or no signal ($\emptyset$)---are given by
\[
\begin{array}{rcl}
\widehat{p}_S(0) &=& \dfrac{\int_0^{p_{Sl}} p_S \,d F_S(p_S)}{F_S(p_{Sl})}, 
\qquad
\widehat{p}_S(1) = \dfrac{\int_{p_{Sh}}^1 p_S \,d F_S(p_S)}{1 - F_S(p_{Sh})}, \\[1.1em]
\widehat{p}_S(\emptyset) &=& 
\dfrac{\widehat{p}_R \int_0^{p_{Sh}} p_S\, dF_S(p_S)
+ (1-\widehat{p}_R)\int_{p_{Sl}}^1 p_S\, dF_S(p_S)}
{\widehat{p}_R F_S(p_{Sh}) + (1-\widehat{p}_R)\big(1-F_S(p_{Sl})\big)}.
\end{array}
\]

If a Responsive Equilibrium exists, receiving $\sigma=1$ leads $R$ to update her belief about the sender's worldview upwards, while receiving $\sigma=0$ leads her to adjust it downwards. 
Note the following important fact: $\widehat{p}_S(\emptyset)$ depends on the worldview of the receiver, whereas $\widehat{p}_S(0)$ and $\widehat{p}_S(1)$ do not. 
The reason is that, in the latter cases, $R$ learns the realization of the signal.
In a Responsive Equilibrium, the following two indifference conditions must hold and determine $p_{Sl}$ and $p_{Sh}$:
\begin{eqnarray}\label{wwIndif1}
C_l &\equiv& -\left| p_{Sl} - \widehat{p}_S(0) \right| - c_S 
+ \int_0^1 \left| \widehat{p}_S(\emptyset) - p_{Sl} \right| \, dF_R(p_R) = 0, \\[0.6em]
C_h &\equiv& -\left| p_{Sh} - \widehat{p}_S(1) \right| - c_S 
+ \int_0^1 \left| \widehat{p}_S(\emptyset) - p_{Sh} \right| \, dF_R(p_R) = 0.
\label{wwIndif2}
\end{eqnarray}

The first condition implies that a sender with worldview $p_{Sl}$ is indifferent between sharing $\sigma=0$ and not sharing her signal, while the second condition implies the same for sharing $\sigma=1$ and not sharing when the sender's worldview is $p_{Sh}$. 
A Responsive Equilibrium is a pair $(p_{Sl}^*,p_{Sh}^*)$
such that both conditions are satisfied simultaneously.
Our next result shows that, if $c_S$ is not too large, then a Responsive Equilibrium 
always exists:

\begin{prop}\label{Prop:Wview} 
Define 
$\bar{c}_S \equiv \min\{\, 1-\xi,\, \mathbb{E}[\,p_S \mid p_S \leq \xi\,] \,\}$.
There exists $\xi \in (0,1)$, determined by $F_S$, such that for all $c_S \in [0,\bar{c}_S)$, an interior Responsive Equilibrium exists with $(p_{Sl}^*,p_{Sh}^*) \in (0,\xi) \times (\xi,1)$.
\end{prop}

If the condition on $c_S$ in the proposition is violated, then an interior equilibrium may not exist. 
In that case, either nobody shares $\sigma=1$, nobody shares $\sigma=0$, or no signals are shared at all. To facilitate the further analysis, we first assume that there is a \textit{single receiver} with a \textit{known worldview} $p_R$.\footnote{Alternatively, we can assume multiple receivers with identical worldview $p_R$.}  This simplifies matters significantly, because we can show that in any interior equilibrium,  it holds that $p_{Sh}^*>\widehat{p}_S^*(\emptyset)>p_{Sl}^*$---see Lemma \ref{lemma:ranking} in Appendix \ref{app:proofReceiver_WW_pR}. Then, the equilibrium conditions \eqref{wwIndif1} and \eqref{wwIndif2} can be written as
\begin{eqnarray}\label{wwIndif3}
\tilde{C}_l &\equiv& -\left(p_{Sl} - \widehat{p}_S(0) \right) - c_S 
+ \left(\widehat{p}_S(\emptyset) - p_{Sl} \right) = 0, \\[0.6em]
\tilde{C}_h &\equiv& -\left(\widehat{p}_S(1)-p_{Sh}\right) - c_S 
+ \left(p_{Sh} -\widehat{p}_S(\emptyset)\right) = 0.
\label{wwIndif4}
\end{eqnarray}
Equipped with these simplified conditions, we now study how sharing costs $c_S$ and the receiver's worldview $p_R$ influence the equilibrium $(p_{Sl}^*,p_{Sh}^*)$. 
We henceforth assume that $c_S \leq \bar{c}_S$, implying that an interior Responsive Equilibrium exists.

To make statements about the comparative statics, we need to guarantee that a \textit{unique} Responsive Equilibrium exists. 
Under Assumption~\ref{assume:Jacobian} below, this is generally the case. 

\begin{assume}
\label{assume:Jacobian}
$\frac{\partial \tilde{C}_l}{\partial p_{Sl}} < 0$, $\frac{\partial \tilde{C}_h}{\partial p_{Sh}} > 0$, and 
$\frac{\partial \tilde{C}_l}{\partial p_{Sl}} \frac{\partial \tilde{C}_h}{\partial p_{Sh}} 
- \frac{\partial \tilde{C}_l}{\partial p_{Sh}} \frac{\partial\tilde{C}_h}{\partial p_{Sl}} < 0$
for all $(p_{Sl},p_{Sh}) \in [0,\xi] \times [\xi,1]$.
\end{assume}

\begin{figure}[t]
  \centering
\includegraphics[width=0.6\textwidth]{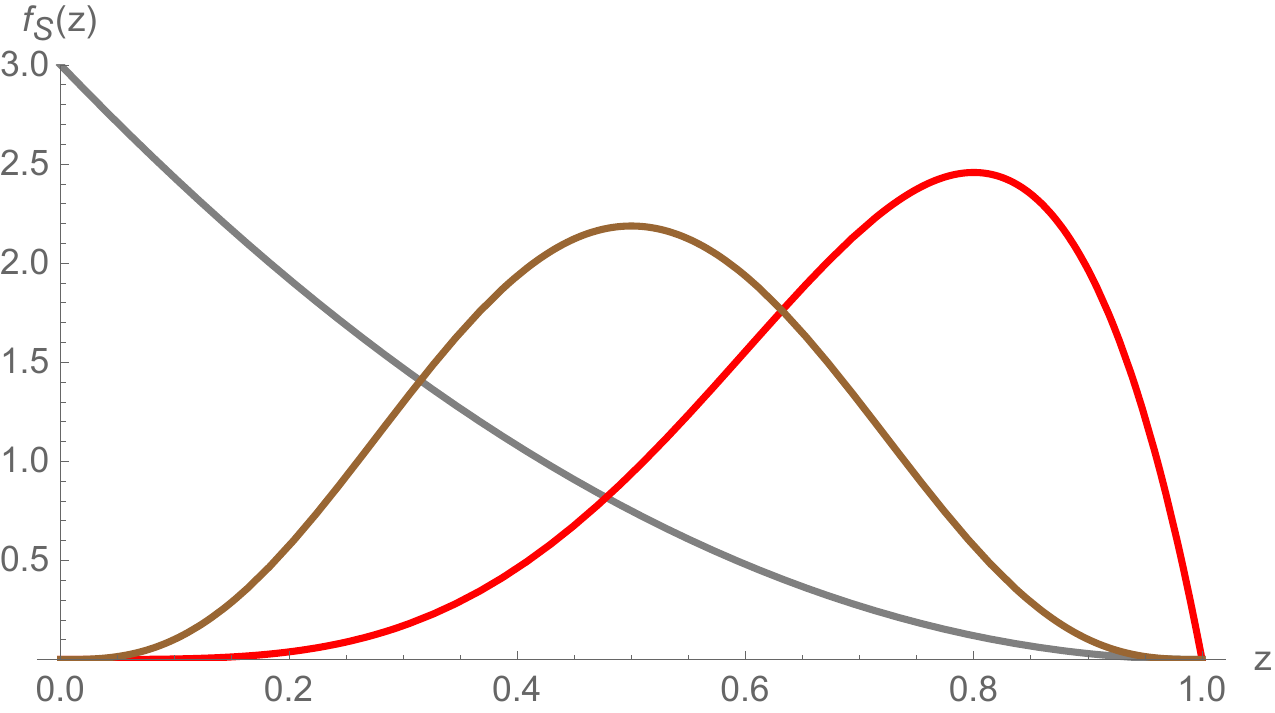}
\caption{Possible shapes of $f_S$ such that Assumption \ref{assume:Jacobian} is satisfied. }\label{Fig:pdfs}
\end{figure}

These assumptions ensure that the Jacobian of the equilibrium conditions $(-\tilde{C}_l,\tilde{C}_h)$ is a P-matrix, in which case uniqueness of the Responsive Equilibrium follows from \cite{GaleNikaido:1965}. 
Note that Assumption~\ref{assume:Jacobian} is not very restrictive. 
Figure~\ref{Fig:pdfs} illustrates several examples of $f_S$ for which Assumption~\ref{assume:Jacobian} holds.\footnote{It is cumbersome to find general conditions on $F_S$ that guarantee Assumption~\ref{assume:Jacobian} without being overly restrictive. 
In Supplementary Appendix~B.1, we show that it generally holds if $F_S$ is uniform on $[\underline{p}_S,\overline{p}_S]\subseteq[0,1]$. 
Moreover, in Supplementary Appendix~B.2 we show that when 
$\frac{f_S(p_S)p_S}{F_S(p_S)} \leq 2$ for $p_S \in [0,\ot]$, 
Assumption~\ref{assume:Jacobian}$\,$ holds in symmetric games, i.e., when $f_S$ is symmetric around $\ot$ and $\widehat{p}_R = \ot$.} 
For the remainder of the analysis, we assume that Assumption~\ref{assume:Jacobian} is satisfied.

Under Assumption \ref{assume:Jacobian}, the comparative statics of the equilibrium thresholds $(p_{Sl}^*,p_{Sh}^*)$ with respect to the receiver's worldview $p_R$ and sharing costs $c_S$ are clear:

\begin{prop}\label{Prop:Receiver_WW_pR}
With a unique receiver with known worldview $p_R$, in any interior equilibrium the following holds: 
$\frac{\partial p_{Sl}^*}{\partial p_R}<0$, $\frac{\partial p_{Sh}^*}{\partial p_R}<0$, $\frac{\partial p_{Sl}^*}{\partial c_S}<0$, and $\frac{\partial p_{Sh}^*}{\partial c_S}>0$.
\end{prop}


The proposition reveals that an \textit{echo chamber effect} emerges endogenously: when the receiver's prior belief about $\omega$ ($p_R$) increases, she receives more signals conforming with her prior, and fewer contradicting her prior. Therefore, to an external observer, the behavior of $S$  may appear to stem from a conformity motive, because she adjusts her strategy toward the receiver's belief. However, it actually stems from moderate sender types, who are wary of being perceived as extreme  when facing a receiver with extreme beliefs. Since these receivers expect  signals to be aligned with their belief, not sharing is a strong indication of opposing beliefs.

Higher sharing costs $c_S$, on the other hand, imply that both $\sigma=1$ and $\sigma=0$ are shared less frequently. Intuitively, when $c_S$ increases, the utility differential between sharing and not sharing needs to increase for sharing to remain worthwhile. A lower (higher) sharing threshold $p_{Sl}^*$ ($p_{Sh}^*$) achieves exactly that: for a sender slightly below (above) the sharing threshold, the utility from sharing increases, while the utility from not sharing remains approximately constant.

\subsection{Multiple Heterogeneous Receivers}
We now discuss what happens when there are multiple receivers with differing worldviews, or when the receiver's worldview is not precisely known. Intuitively, the comparative statics with respect to $c_S$ are qualitatively unchanged to those in Proposition \ref{Prop:Receiver_WW_pR}. However, when the distribution of $p_R$ changes, things are not as clear unless we make further assumptions. Our last result of this section is a direct  corollary of Proposition \ref{Prop:Receiver_WW_pR} and shows that the described echo chamber effect remains valid if receivers are ideologically sufficiently homogeneous and if there is a population wide ideological drift:

\begin{corollary}
Consider two distinct distributions of receiver ideologies, 
$F_R(p_R)$ with support $[\underline{\alpha},\overline{\alpha}]\subseteq [0,1]$ and 
$\hat{F}_R(p_R)$ with support $[\underline{\beta},\overline{\beta}]\subseteq [0,1]$. Assume we can rank the two distributions by first-order stochastic dominance: $\hat{F}_R(p_R)\leq F_R(p_R)$. If the receiver population under each distribution is sufficiently homogeneous---both $|\overline{\alpha}-\underline{\alpha}|$ and $|\overline{\beta}-\underline{\beta}|$ are sufficiently small---then $p_{Sl}^{*}[\hat{F}_R]<p^*_{Sl}[F_R]$ and $p^*_{Sh}[\hat{F}_R]<p^*_{Sh}[F_R]$ in any interior equilibrium.
\label{prop:multi}
\end{corollary}

This result directly follows from Proposition \ref{Prop:Receiver_WW_pR}. Intuitively, starting at $F_R$ and moving towards $\hat{F}_R$, we  successively increase the ideology $p_R$ of some receivers. Because under both distributions audiences are sufficiently homogeneous, the ranking $p_{Sh}^*>\widehat{p}_S^*(\emptyset)>p_{Sl}^*$ holds for any $p_R$, and thus each of these increases leads to lower equilibrium thresholds by the echo chamber effect discussed above. Hence, 
the sum of all of these shifts needs to lead to  lower sharing thresholds $p_{Sl}$ and $p_{Sh}$ as well. 

With a sufficiently heterogeneous audience, it will often be the case  that $p_{Sh}^*>\widehat{p}_S^*(\emptyset)>p_{Sl}^*$ does not hold anymore for some (extreme) receiver types, implying this logic is not valid anymore. However, this does not imply that our comparative static result does not hold anymore. In Supplementary Appendix B.3 we show in an example that for a maximally polarized audience where $\widehat{p}_R\in\{0,1\}$, $p_{Sh}^*>\widehat{p}_S^*(\emptyset)>p_{Sl}^*$ is always violated for one of the receiver types, but the comparative static in Corollary \ref{prop:multi} still holds.

\subsection{The Quality of Shared Information}

We now study which situations are conducive to the spread of fake news.  
As in Section \ref{sec:ability}, we use the share of improper signals relative to all signals shared as a measure of the quality of information. This fraction equals

\begin{eqnarray}\label{Eq:ShareFake}
    \gamma=\frac{q(1-\beta)F_S(p_{Sl})+q\beta(1-F_S(p_{Sh}))}{\mathbb{P}(\sigma=0)F_S(p_{Sl})+\mathbb{P}(\sigma=1)(1-F_S(p_{Sh}))},
\end{eqnarray}
\normalsize

\noindent
where $\mathbb{P}(\sigma=1)=q\beta+(1-q)\left[p_T \eta+(1-p_T)(1-\eta)\right]$ 
and $\mathbb{P}(\sigma=0)=1-\mathbb{P}(\sigma=1)$
are the true probabilities of receiving signals with the different realizations. As in Section \ref{sec:ability}$, \gamma$ can be interpreted as an inverse measure of the quality of information after sharing. The higher is $\gamma$, the lower the quality of information. Furthermore, whenever $\gamma>q$, the quality of information deteriorates after sharing. 

For simplicity, we maintain our assumption of one receiver with known worldview $p_R$.\footnote{We can again generalize our results to receiver populations that are sufficiently homogeneous, using analogous arguments as for Corollary \ref{prop:multi}.} We now state our first result on the spread of fake news when social image concerns regarding one's worldview are relevant. 
\begin{prop}\label{Prop:WorldviewFakeGeneral}
Define $\hat{\beta}\equiv p_T \eta+(1-p_T)(1-\eta)$. If $\beta=\hat{\beta}$, then the quality of information does not change after sharing, $\gamma=q$. Moreover, if $\beta>\hat{\beta}$ ($\beta<\hat{\beta}$), then $\gamma$
increases (decreases) in $p_R$.
\end{prop}

$\hat{\beta}$ denotes the bias of improper signals towards 1 at which signal realization $\sigma=1$ is equally likely among proper and improper signals. Since $S$ does not condition on veracity when making her sharing decision, but only on signal realization $\sigma$, the quality of information conditional on $\sigma$ does not change after sharing. If $\beta>\hat{\beta}$, $\sigma=1$ is more likely among improper signals than among proper signals, while the opposite holds when $\beta<\hat{\beta}$. If a realization that is more likely to be fake than the other 
is relayed by the sender more frequently,
 this will deteriorate information quality. 
 
 Proposition \ref{Prop:Receiver_WW_pR} shows that there is a positive correlation between the worldview of the receiver and the 
 amount of signals she receives that concur with her worldview. Thus, when social image concerns revolve around worldview, the problem of fake news sharing is especially relevant when the worldview of the receiver is aligned with the bias of the fake news shared.  This is in contrast to the ability motive, where the problem of fake news sharing was especially relevant when fake news were biased against the belief of the receiver and surprising signals were shared disproportionally.

The next proposition states when sharing leads to a deterioration of the quality of information:

\begin{prop}\label{Prop:WvInfDet}
In any interior equilibrium, the following holds:
\[
\text{Sign}\left[\gamma-q\right]=\text{Sign}\left[(\hat{\beta}-\beta)\left[F_S(p_{Sl}^*)-(1-F_S(p_{Sh}^*))\right]\right].
\]
\end{prop}

From Proposition \ref{Prop:WvInfDet} it is clear that information deteriorates through sharing ($\gamma>q$) if fake signals are biased towards $x\in\{0,1\}$ and $\sigma=x$ is shared more often than $\sigma=1-x$.

While the receiver's worldview $p_{R}$ is an important determinant of the sign of $F_S(p_{Sl}^*)-(1-F_S(p_{Sh}^*))$, as also shown in Proposition \ref{Prop:Receiver_WW_pR}, the sender's strategy depends on it only through $\widehat{p}_R$, the receiver's belief that the sender received a signal $\sigma=1$. If fake news are prevalent and strongly biased, $\widehat{p}_R\approx q \beta$. For example, if $R$ expects 
$\sigma=1$ because $q\beta$ is very high, moderate types are reluctant not to share such signals, as not sharing is a strong signal of having an opposing worldview. Because of this echo chamber effect,  for high $q$ the quality of information after sharing is likely  to be lower than before sharing. 

\begin{figure}[t]
  \centering
\includegraphics[width=0.6\textwidth]{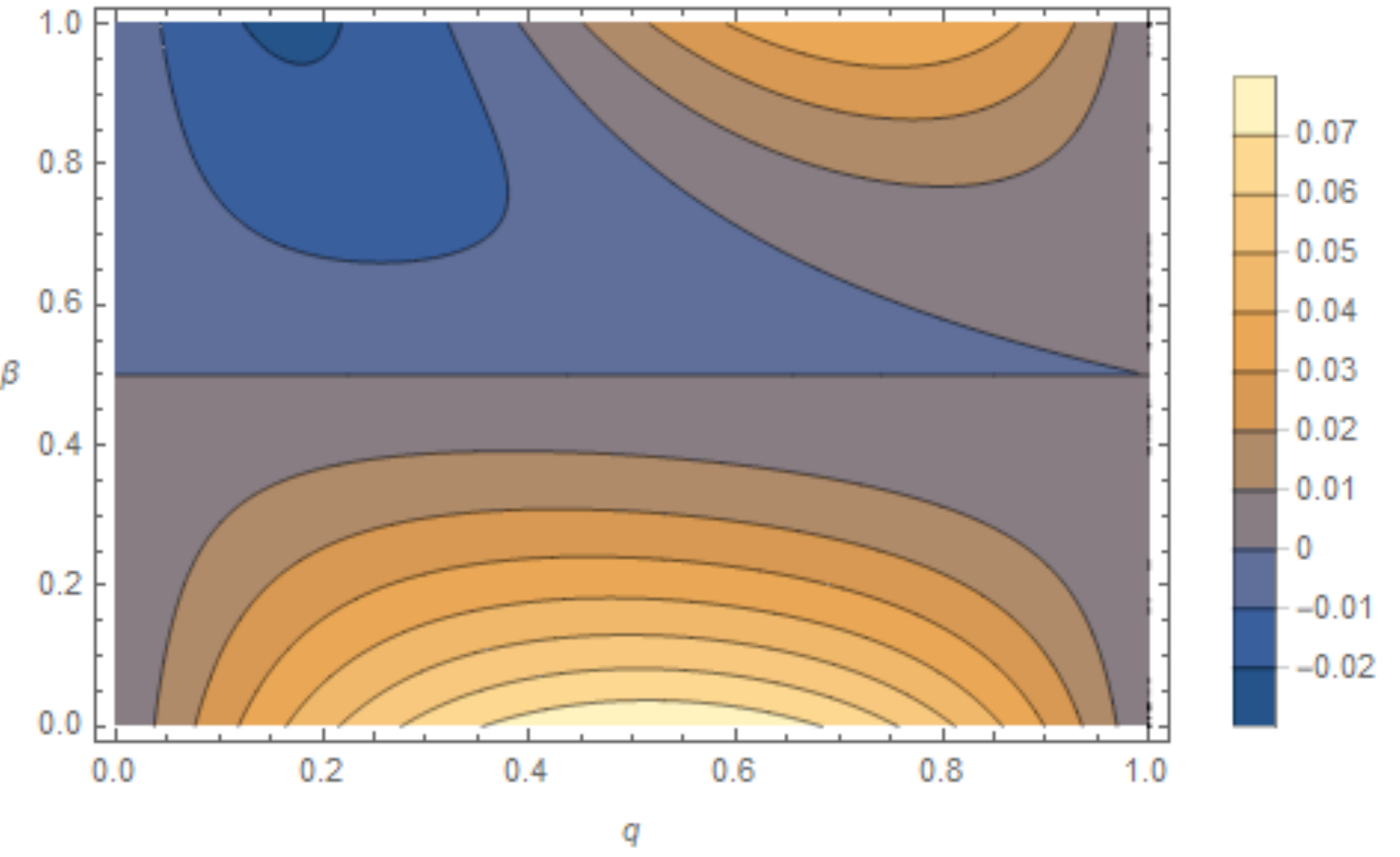}
\caption{Quality of signals after relative to before sharing ($\gamma-q$) as a function of $q$ and $\beta$, given $\eta=\frac{9}{10}$, $p_R=\frac{1}{10}$ and $p_T=\frac{1}{2}$. Positive values (yellow tones) denote a deterioration of information quality after sharing. }\label{Fig:FakeSpreadWv}
\end{figure}

To illustrate this effect, in Figure \ref{Fig:FakeSpreadWv} we plot $\gamma-q$ assuming that $p_S$ follows a uniform distribution on $[0,1]$.
For low values of $q$ we see that the quality of information deteriorates for $\beta<\ot=\hat{\beta}$. For high values of $q$, on the other hand, information quality deteriorates for any $\beta$ that is not too close to $\frac{1}{2}$. 
For high enough $q$, the belief of the receiver about $\omega$  is not very relevant for her belief about the signal realization. She will expect more $\sigma=1$ when $\beta>\ot$ and more $\sigma=0$  else. Social image concerns about one's worldview together with a high prevalence of fake news thus provide a fertile ground for excessive fake news sharing. 

\subsection{Receiver Welfare}

Finally, we turn to receiver welfare. First of all, signals that align with her prior are not relevant to the receiver, and as we showed, relevant signals are shared relatively infrequently with her. At the same time, senders do not ``filter'' signals conditional on their headline, thus the quality of information before and after sharing \emph{conditional on $\sigma$} is the same. This also implies that increasing sharing cost $c_S$, which discourages sharing on the margin, will weakly decrease receiver welfare as less signals are shared which are potentially informative. This is in contrast to ability signaling, where sharing costs differentially discourage low-ability senders from sharing and thus increase the quality of information shared conditional on $\sigma$, potentially increasing welfare of low-ability types.

While senders do not selectively share fake or proper signals, they do tailor their strategy to the receiver's worldview, such that receivers with extreme views receive more signals aligned with their worldview than moderates. In our model, rational receivers also update their belief from not receiving a signal. If this were not the case, for example because of selection neglect, we should see an increase in belief polarization as in \cite{BowenEtAl:2023} because of the worldview motive. Our worldview model thus provides a micro-foundation for selective sharing in low-attention information environments. 

\section{Combining Ability and Worldview Motives}
\label{sec:mixed_motives}

The analyses in Sections \ref{sec:ability} and \ref{sec:worldview} deliberately separated the two social image motives to provide a rich analysis of when fake news sharing is especially problematic under each motive. In this section we combine the two motives in a simplified framework to study how they interact. Assume that a fraction $\mu \in (0,1)$ of high-ability senders cares about looking competent, as in Section \ref{sec:ability} while the remaining high-ability senders and all low-ability senders care about signaling their worldview, as in Section \ref{sec:worldview}. We characterize equilibria that illustrate how different types of signals serve to signal different individual characteristics and how this can foster fake news sharing by high-ability types.

In contrast to the main model, we now assume that not just ability but also worldview is binary, $p_i\in\{p_L,p_H\}$, $i=\{R,S\}$. In addition, we assume symmetry in that a) both sender and receiver are equally likely to be of high or low worldview type and b) these types are equidistant from one half, $p_L=1-p_H$ and $0 < p_L < 1/2 < p_H < 1$. In addition, both state realizations are equally likely, $p_T=\frac{1}{2}$, and fake signals are unbiased, $\beta = \frac{1}{2}$. For simplicity, also assume that all receivers are of the high-ability type and verify \emph{all} signals.\footnote{We do this to prevent the analysis from becoming overly long. Similar results can be obtained for $\lambda_R$ and $\beta$ as free parameters and under the original fact-checking assumption introduced in Section \ref{sec:model}. Details are available upon request.}

Under these assumptions, the model is fully symmetric: the population of senders and receivers is balanced across worldviews, and both proper and fake signals are equally likely to carry either content, $\sigma=1$ or $\sigma=0$. While this symmetry does not serve to study the severity of fake news sharing generally, which we learned in previous sections arises typically when either fake signals are sufficiently biased or sender and receiver worldviews are extreme and either conform or conflict, it serves to succinctly illustrate an additional channel through which fake news sharing becomes problematic. It highlights how receivers infer multiple characteristics from one sharing decision, and how different types of signals are thus useful for signaling different characteristics. 


As in Section \ref{sec:worldview} we focus on equilibria with natural language and thus $\sigma=1$ should weakly increase $R$'s estimate of $S$'s worldview (towards $p_H$) while $\sigma=0$ should weakly decrease it (towards $p_L$). We thus define \emph{congruent} signals as $\sigma=0$ for $p_S=p_L$ types and $\sigma=1$ for $p_S=p_H$ types. The following proposition states our result pertaining to this framework\footnote{The proof can be found in Supplementary Appendix C.}:

\begin{prop}
\label{prop:mixed_motives}
Consider the symmetric set-up described above. There exist sharing cost thresholds
\[
0 < c_S^{ii}  < c_S^{i}
\]
such that the following holds.

\begin{enumerate}[label=(\roman*), leftmargin=1.25cm]
  \item If $c_S \le c_S^{i}$, there exists an equilibrium in which
worldview-motivated high-ability senders do not check congruent signals and
share all congruent signals, both proper and fake.
    \item If $c_S \geq c_S^{ii}$ but sufficiently small, as well as $\lambda_S\geq\tfrac15$, $10 \mu\geq 5+\frac{1}{\lambda_S}$ and $q\leq\frac{1+5 \lambda_S-10 \mu \lambda_S}{3 \lambda_S-9 \mu \lambda_S}$, there exists an equilibrium in which worldview-motivated high-ability senders check the veracity status of congruent signals and share \emph{only fake} congruent signals and \emph{do not} share proper congruent and all incongruent signals.
\end{enumerate}

In both cases, low-ability (worldview-motivated) senders share congruent content only, ability-motivated high-ability senders check all signals and share all proper signals and no fake signals.
\end{prop}

As in Section \ref{sec:ability}, ability-motivated high-ability types filter signals and thus improve the quality of information after sharing. As in Section \ref{sec:worldview}, worldview-motivated low-ability types do not filter information by quality, only filtering on content. In addition, we now find that the presence of sufficiently many ability-motivated high-ability types using proper signals to signal ability increases the incentives of worldview-motivated high-ability types to share \emph{improper} signals. In the extreme case, the proposition shows that high-ability types pursuing worldview image concerns might only share fake signals as sharing costs increase, thus deteriorating the quality of information after sharing.\footnote{Note that for low enough sharing costs also an equilibrium exists, where ability-motivated types never share, while worldview-motivated types only share congruent signals and perfectly reveal their type. Thus, ability-signaling through sharing is shut down in this equilibrium. The reverse is also possible for low enough sharing costs: ability-motivated senders share only proper signals while worldview-motivated senders do not share. Off-equilibrium beliefs after observing fake signals are equal to the prior.} 




\section{Discussion}
\label{sec:discussion}
Our analysis shows that both social image motives can lead to disproportionate sharing of fake news. The motives, however, generate distinct sharing patterns, both in who shares fake news and in how its content relates to receivers' beliefs. Our empirical predictions build on the idea that different social image motives are more salient in different environments. We therefore focus on the benchmark models in Sections \ref{sec:ability} and \ref{sec:worldview}, which isolate the implications of each motive. The simplified mixed-motive framework in Section \ref{sec:mixed_motives} is primarily meant to illustrate how the two motives can interact rather than to deliver general empirical predictions. We now formulate the corresponding hypotheses and compare them with existing evidence.


We begin by formulating hypotheses about sharing behaviour conditional on the different motives to share information. 
With an \textit{ability signaling motive}, the following hypotheses follow from our analysis:
\setcounter{hypothesis}{0}
\begin{hypothesis}\label{hypo:A}
Consider a salient ability signaling motive.
\begin{enumerate}[label=(\alph*)]
\item \label{hypA:itemb}Fake news are shared predominantly by users with a low ability.
\item \label{hypA:itemc}Fake news are typically novel/surprising for the intended audience. 
\item \label{hypA:itemd}Fake news are mostly shared by users with opposing ideology as the intended receiver(s). 
\end{enumerate}
\end{hypothesis}
\setcounter{hypothesis}{22}
With a \textit{worldview signaling motive}, the hypotheses derived from our model are  distinct:
\begin{hypothesis}\label{hypo:W}
Consider a salient worldview signaling motive.
\begin{enumerate}[label=(\alph*)]
    \item\label{hypW:itema} There is little to no engagement with the shared news, and the expected quality of the news is irrelevant for the sharing  decision.
\item\label{hypW:itemb} Moderates tend to share no information, whereas those on the extremes of the ideological spectrum share information in line with their belief, both fake and proper. 
\item\label{hypW:itemc} Fake news are mostly shared by users with similar ideology as the intended receiver(s).
\item\label{hypW:itemd} The propensity to share fake news is orthogonal to ability.
\end{enumerate}
\end{hypothesis}

To distinguish between the two motives empirically, we consider how their salience varies across social networks. Closed networks, such as Facebook and Instagram, consist largely of pre-existing social ties, whereas audiences on open networks such as Twitter/$\mathbb{X}$ tend to be more heterogeneous. Audience homogeneity and familiarity facilitate identity-driven self-presentation and self-disclosure \citep{bazarova2014self, vanDerDoes2022strategic, altman1973social, greene2006self}. We therefore expect worldview signaling to be relatively more salient in closed networks. Consistent with this argument, \citet{marwick2011tweet} document greater self-censorship on Twitter/$\mathbb{X}$, while the platform is used disproportionately for disseminating scholarly work \citep{erdt2016altmetrics, enkhbayar2020much}, which is more consistent with ability signaling.

We first consider open or public networks, where we expect ability signaling to be relatively salient. \citet{GurievHenryMarquisZhuravskaya:2023} estimate a structural model of news sharing on Twitter/$\mathbb{X}$ and find ability signaling to be the empirically most important motive behind sharing. This is consistent with the premise underlying Hypothesis \ref{hypo:A}. Further evidence comes from \citet{VosoughiRoyAral:2018}. Using Twitter/$\mathbb{X}$ data, they show that false news is more novel than factual news and that users who spread false news have significantly fewer followers. If follower counts are positively related to ability, both of these are consistent with Hypothesis \ref{hypo:A} as well.\footnote{Greater network centrality has been found to correlate positively with educational and academic achievement; see, for example, \citet{park2020trust}, \citet{williams2019linking}, and \citet{VIGNERY2020101499}.}


Turning now to evidence from closed networks, where we expect worldview signaling to be salient. \cite{SundarEtAl:2025} analyze over 35 million Facebook posts containing URLs shared between 2017 and 2020. They find that around 75\% are ``\textit{shares without clicks},'' and thus most senders do not engage with the shared content themselves. Differences in the propensity of sharing false news without previously clicking on the URL between individuals of different groups are found to be due to differences in the number of false URLs originating from news domains frequented by those groups. This is consistent with Part \ref{hypW:itema} of Hypothesis \ref{hypo:W}. Moreover, copartisan political content receives more shares without clicks, with partisans engaging in it more than politically neutral users. \cite{GuessEtAl:2019} and \cite{GuessEtAl:2023} come to a similar conclusion, also using data from Facebook. This is consistent with Part \ref{hypW:itemb} of Hypothesis \ref{hypo:W}.

In surveys run on MTurk, \cite{PennycookEtAl:2021} find further evidence for this hypothesis, asking individuals if they intend to share a certain piece of information: ``\textit{Whether the headline was politically concordant or discordant had a significantly larger effect on sharing intentions [\dots] than whether the headline was true}.'' Moreover, ``\textit{participants were more than twice as likely to consider sharing false but politically concordant headlines as they were to rate such headlines as accurate.}''  It appears that \cite{PennycookEtAl:2021} did not clearly specify on which social network participants were to share these headlines. Therefore, it seems likely that a majority of the subjects had a closed network in mind, as the by far largest network in the U.S. at the time was Facebook, with more than three times as many users as Twitter.\footnote{See \url{https://www.pewresearch.org/short-reads/2019/04/10/share-of-u-s-adults-using-social media-including-facebook-is-mostly-unchanged-since-2018/}.} This suggests worldview signaling was more salient at the time of the survey, and their results are consistent with Hypothesis \ref{hypo:W}. However, they also show that by explicitly increasing the salience of the ability motive through accuracy nudges,  the willingness to share misinformation decreased significantly among survey respondents. An accompanying Twitter intervention had the same result. 

The effectiveness of such accuracy nudges can be explained through different mechanisms.
 \cite{stein2024partisan} show that accuracy beliefs are resistant to accuracy incentives. Instead, we argue that the salience of ability signaling increases through accuracy nudges. In a follow-up study to \cite{PennycookEtAl:2021}, \cite{lin2023thinking} show exactly that: these behavioral changes are achieved by increasing ``\textit{the weight participants put on accuracy while deliberating}.'' \cite{GurievHenryMarquisZhuravskaya:2023} study a similar accuracy nudge intervention (among others) and show through structural estimation that it indeed works by shifting weight toward reputational (ability) concerns in the sharing decision.

In line with these findings, our model suggests that in situations where fake news is shared disproportionately under a worldview motive, nudging towards accuracy---and thus an ability motive---may increase the quality of shared information. This will be the case if the bias of the fake news aligns with the receiver's beliefs; in such a situation, under a worldview motive fake news sharing can be significant (Hypothesis \ref{hypo:W}), whereas it tends to be more limited under an ability motive (Hypothesis \ref{hypo:A}). However, in  situations where fake news tends to be surprising, worldview signaling may lead to higher overall quality of shared information than ability signaling.
However, note that this is driven to a large extent by fake \textit{decision-irrelevant} signals---that is, signals that do not change the receiver's action irrespective of their veracity---which are shared under a worldview motive but not under an ability motive. The quality of information \emph{conditional on relevance} should generally increase under such a nudge. Therefore, even though  accuracy nudges may lead to greater spread of fake information, the receiver's decision quality should improve.

Another implication of our analysis is that one should be cautious to use  institutional fact-checking, because such efforts can have nuanced effects. As long as the ability motive is salient, incentives will lead senders to fact-check. Hence, institutional fact-checking can improve the quality of information by increasing the number of ``high-ability'' senders. But if institutional fact-checking is available to everyone, it may shut down ability signaling altogether. This way, institutional fact-checking may increase the salience of worldview signaling, and thus potentially decrease the quality of shared information.

Finally, our analysis also highlights the consequences of different social image concerns for the type of information individuals receive on social media. Under an ability motive, social media users predominantly receive decision-relevant information---that is, information that challenges their prior beliefs. This differs under a worldview motive, where audiences receive information likely aligned with their worldview, and information that challenges it is less prevalent. While under both motives senders tailor their strategy to their audience, this tendency is especially pronounced under the worldview motive, where individuals with extreme worldviews also receive the most unbalanced news diets through their networks. So far, the empirical literature has typically focused on sender beliefs and behavior. However, studying the beliefs and polarization levels of social media audiences under different information-sharing motives also seems a  fruitful avenue for future research. Apart from the information contained in the news shared, particularly under the worldview motive, individuals also receive information about others' worldviews and beliefs. The consequences of this for belief polarization and network structure have so far received relatively little attention in the literature.

\appendix

\section{Mathematical Appendix}

\subsection{Proof of Proposition \ref{prop:EQ}}
Throughout this proof, let
$\tilde q^S\equiv\tilde q^S(0,p_S,U)$
denote the belief of a low-ability sender that a surprising signal is fake.
We first prove an important lemma:

\begin{lem}\label{lem:delta-mono}
$\Delta$ is strictly decreasing in $\kappa(0,L,\tilde q)$.
\end{lem}

\begin{proof}
Both $\pi_{0P}$ and $\pi_{0U}$ are strictly decreasing in
$\kappa(0,L,\tilde q)$, so $u_{0U} = \lambda_R(1 - \tilde q^S)\pi_{0P} +
(1-\lambda_R)\pi_{0U} - c_S$ is strictly decreasing in it as well. Silence
pools more low ability senders as $\kappa(0,L,\tilde q)$ rises, so
$u_\emptyset = \pi_\emptyset$ is strictly increasing. Hence
$\Delta = u_{0U} - u_\emptyset$ is strictly decreasing.
\end{proof}

Now define the probability estimate of player $i\in\{S,R\}$ that the signal generation process leads to  a surprising signal, $z_0^i$, and a proper surprising signal, $z_{0\mathcal{P}}^i$, by
\[
\begin{array}{rcl}
z_0^i&=&q (1-\beta)+(1-q)\left(p_i(1-\eta)+(1-p_i)\eta\right),\\  
z_{0\mathcal{P}}^i&=&(1-q)\left(p_i(1-\eta)+(1-p_i)\eta\right).
\end{array}
\]
The probability of receiving a fake and surprising signal follows from these two probabilities and is $z_{0\mathcal{F}}^i=z_{0}^i-z_{0\mathcal{P}}^i=q(1-\beta)$.
Given our above-defined notation, and assuming the claimed equilibrium probabilities $\kappa(1,H,\tilde{q}^S)=\kappa(0,H,1)=\kappa(1,L,\tilde{q}^S)=0$ and $\kappa(0,H,0)=1$, we can now define the relevant beliefs held by receivers about sender's ability $\theta_S$.
If a surprising signal is relayed, there are three different beliefs to keep track of. A high ability receiver will check the signal's veracity and thus holds belief $\pi_{0\mathcal{F}}=0$ if the signal is fake. If the signal is proper, her belief is
\[
\pi_{0\mathcal{P}}=\frac{\lambda_S z_{0\mathcal{P}}^R}{\lambda_S z_{0\mathcal{P}}^R+(1-\lambda_S)z_{0\mathcal{P}}^R\kL}=\frac{\lambda_S}{\lambda_S+(1-\lambda_S)\kappa(0,L,\tilde{q})}.
\]
A low ability receiver does not know the signal's veracity and hence holds belief
\[
\pi_{0\mathcal{U}}=\frac{\lambda_S z_{0\mathcal{P}}^R}{\lambda_S z_{0\mathcal{P}}^R+(1-\lambda_S)z_0^R\kL}.
\]
Note that $\pi_{0\mathcal{P}}>\pi_{0\mathcal{U}}$ because $z_0^R>z_{0\mathcal{P}}^R$. Moreover, $\pi_{0\mathcal{P}}>\lambda_S$, because $\kL<1$.
If no signal is shared, the belief is
\[
\pi_\emptyset=\frac{\lambda_S(1-z_{0\mathcal{P}}^R)}{\lambda_S(1-z_{0\mathcal{P}}^R)+(1-\lambda_S)\left[(1-z_0^R+z_0^R(1-\kL)\right]}.
\]
Finally, the belief a low ability sender holds regarding the veracity of a surprising signal she holds is
\begin{equation}
\tilde{q}^S=\frac{z_{0\mathcal{F}}^S}{z_{0\mathcal{F}}^S+z_{0\mathcal{P}}^S}=\frac{z_{0\mathcal{F}}^S}{z_{0}^S}.
\label{eq:qtilde}
\end{equation}
Thus, $
\partial \tilde{q}^S/\partial q>0$, because
\[
\dfrac{\partial \tilde{q}^S}{\partial q}=\dfrac{(1-\beta) \left((1-2 \eta) p_S+\eta \right)}{\left(p_S (2 \eta-1)(1-q)+q (\beta +\eta -1)-\eta\right)^2}>0\Leftrightarrow (1-2 \eta) p_S+\eta >0.
\]
The last inequality follows from the LHS being decreasing in $p_S$, and when $p_S=1$ we have $1-2\eta+\eta=1-\eta>0$. We need these different beliefs to define $S$'s expected utility from sharing a given signal. If the sender knows a surprising signal to be proper, she receives expected utility from sharing it equal to $u_{0\mathcal{P}}$ while sharing a signal that is known to be fake yields $u_{0\mathcal{F}}$ with
$
u_{0\mathcal{P}}=\lambda_R \pi_{0\mathcal{P}}+(1-\lambda_R)\pi_{0\mathcal{U}}-c_S$ and $
u_{0\mathcal{F}}=(1-\lambda_R)\pi_{0\mathcal{U}}-c_S$. If the signal's veracity is not known, sharing a surprising signal yields an expected utility of
\[
\begin{array}{rcl}
u_{0\mathcal{U}}&=&\lambda_R\left(\tilde{q}^S\cdot 0+(1-\tilde{q}^S)\pi_{0\mathcal{P}}\right)+(1-\lambda_R)\pi_{0\mathcal{U}}-c_S\\
&=&\lambda_R(1-\tilde{q}^S)\pi_{0\mathcal{P}}+(1-\lambda_R)\pi_{0\mathcal{U}}-c_S.
\end{array}
\]
Not sharing a signal implies no sharing cost $c_S$, and hence yields expected utility $u_{\emptyset}=\pi_{\emptyset}$. Finally, off-equilibrium sharing of a non-surprising signal yields
$u_1^D=\pi_1-c_S$.

Define \begin{equation}
\label{EQ:Delta}
\Delta\equiv u_{0\mathcal{U}}-u_\emptyset.
\end{equation}
In equilibrium, it must hold that the low ability sender is indifferent between sharing and keeping  her signal, $\Delta=0$. Moreover, she must weakly prefer not to share a non-surprising signal to deviating and sharing it, $u_\emptyset\geq u_1^D$. Hence, we need $u_{0\mathcal{U}}=u_\emptyset\geq u_1^D$.

First assume $\kL=1$.
\begin{eqnarray*}
\left.\Delta\right|_{\kL=1}&=&\lambda_R(1-\tilde{q}^S)\lambda_S+(1-\lambda_R)\frac{\lambda_S z^R_{0\mathcal{P}}}{\lambda_S z^R_{0\mathcal{P}}+(1-\lambda_S)z^R_0}-c_S\\
&&-\frac{\lambda_S(1-z^R_{0\mathcal{P}})}{\lambda_S(1-z^R_{0\mathcal{P}})+(1-\lambda_S)(1-z^R_0)}
\end{eqnarray*}
This expression is decreasing in both $c_S$ and $\tilde{q}^S$. 
We have
\[
\begin{array}{rcl}
\left.\Delta\right|_{\kL=1\wedge c_S=\tilde{q}^S=0}&=&\lambda_R\lambda_S+(1-\lambda_R)\dfrac{\lambda_S z^R_{0\mathcal{P}}}{\lambda_S z^R_{0\mathcal{P}}+(1-\lambda_S)z^R_0}\\
&&-\dfrac{\lambda_S(1-z^R_{0\mathcal{P}})}{\lambda_S(1-z^R_{0\mathcal{P}})+(1-\lambda_S)(1-z^R_0)}<0\\
\\
\Leftrightarrow  \lambda_R+(1-\lambda_R)\dfrac{z^R_{0\mathcal{P}}}{\lambda_S z^R_{0\mathcal{P}}+(1-\lambda_S)z^R_0}&<&\dfrac{(1-z^R_{0\mathcal{P}})}{\lambda_S(1-z^R_{0\mathcal{P}})+(1-\lambda_S)(1-z^R_0)}
\end{array}
\]
The LHS increases in $z^R_{0\mcp}$, while the RHS decreases in it. The greatest possible value it can take is $z^R_0$.
Then the LHS becomes 1, and so does the RHS. But this implies that for any $z^R_{0\mcp}<z^R_0$, $\left.\Delta\right|_{\kL=1\wedge c_S=\tilde{q}^S=0}<0$ holds, and thus also $\left.\Delta\right|_{\kL=1}<0$ holds true. Hence, even if sharing is costless, the low ability sender will keep some signal to herself.

Next assume $\kL=0$. Then
\begin{equation}
\label{eq:Delta0}
\left.\Delta\right|_{\kL=0}=\lambda_R(1-\tilde{q}^S)+(1-\lambda_R)-c_S-\frac{\lambda_S(1-z^R_{0\mathcal{P}})}{\lambda_S(1-z^R_{0\mathcal{P}})+(1-\lambda_S)}
\end{equation}
Our goal is to show that this is positive when $c_S$ is sufficiently small, hence let $c_S=0$. Then
\[
\begin{array}{rcl}
\left.\Delta\right|_{\kL=0\wedge c_S=0}&=&\lambda_R(1-\tilde{q}^S)+(1-\lambda_R)-\dfrac{\lambda_S(1-z^R_{0\mathcal{P}})}{\lambda_S(1-z^R_{0\mathcal{P}})+(1-\lambda_S)}>0\\
\Leftrightarrow&&1-\dfrac{\lambda_S(1-z^R_{0\mathcal{P}})}{\lambda_S(1-z^R_{0\mathcal{P}})+(1-\lambda_S)}>\lambda_R\tilde{q}^S
\Leftrightarrow\dfrac{1}{\lambda_R}\dfrac{1-\lambda_S}{\lambda_S(1-z^R_{0\mathcal{P}})+(1-\lambda_S)}>\tilde{q}^S
\end{array}
\]
The LHS increases in $z_{0\mcp}^R$, while the  RHS is independent of  it. Hence, let $z_{0\mcp}^R=0$. Then it must hold that 
$\frac{1}{\lambda_R}(1-\lambda_S)>
 \tilde{q}^S$. Note that because $\lambda_S+\lambda_R>0$  we have $\frac{1}{\lambda_R}(1-\lambda_S)>1$. Moreover, because   $\tilde{q}^S<1$, the inequality  is satisfied. This implies that as long as $c_S$ is small enough, $\left.\Delta\right|_{\kL=0}>0$. However, when $c_S$ becomes large, this changes. Hence, there exists $\overline{c}_S$ such that if $c_S<\overline{c}_S$, then $\left.\Delta\right|_{\kL=0}>0$. Because for all $c_S$ it holds that $\left.\Delta\right|_{\kL=1}<0$, and because $\Delta$ is a continuous function of $\kL$, it follows from the intermediate value theorem that there exists $\kappa^*(0,L,\tilde{q})\in(0,1)$ such that $\Delta(\kappa^*(0,L,\tilde{q}))=0$.
Now assume $c_S>0$. To prove existence of the equilibrium discussed in the proposition, we need to show that \eqref{eq:Delta0} is positive. Note that $q$ enters \eqref{eq:Delta0} through $\tilde{q}^S$, and $\tilde{q}^S$ increases in $q$. It also enters through $z_{0\mcp}^R$ which decreases in $q$, while \eqref{eq:Delta0} increases in $z_{0\mcp}^R$. Therefore, when $\left.\Delta\right|_{\kL=0}>0$ for some $q'$, then the same is true for all $q<q'$. But does such $q'$ always exist?
When $q\rightarrow 0$, \eqref{eq:Delta0} becomes $
\left.\Delta\right|_{\kL=0\wedge q=0}=1-c_S-\frac{\lambda_S(1-z^R_{0\mathcal{P}})}{\lambda_S(1-z^R_{0\mathcal{P}})+(1-\lambda_S)}>0$.
As long as $
c_S<c_L\equiv\frac{1-\lambda_S}{1-\lambda_S z^R_{0\mcp}}\in(0,1)$, 
this is true. This implies that for all $c_S\in[0,c_L)$ there exists $\bar{q}(c_S)$ such that  if $q\leq \bar{q}(c_S)$, then the equilibrium exists. Moreover, our above analysis reveals that $\bar{q}(0)=1$ and $\bar{q}(c_L)=0$. That $\bar{q}(c_S)$ decreases in $c_S$ follows from
\[
\frac{\partial \bar{q}(c_s)}{\partial c_S}=-\frac{\frac{\partial \left.\Delta\right|_{\kL=0}}{\partial c_S}}{\frac{\partial \left.\Delta\right|_{\kL=0}}{\partial q}}=-\frac{-1}{-\lambda_S \frac{\partial \tilde{q}^S}{\partial q}}=-\frac{1}{\lambda_S \frac{\partial \tilde{q}^S}{\partial q}}<0,
\]
because $\frac{\partial \tilde{q}^S}{\partial q}>0$ (see \eqref{eq:qtilde}).
When the low ability sender is indifferent between sharing and not sharing the signal, the high ability sender has a strict incentive to share a proper and surprising signal, because $u_{0\mcp}>u_{0\mcu}$. Moreover, because $u_{0\mcf}<u_{0\mcu}=u_{\emptyset}$, the high ability sender also keeps a fake but surprising signal to herself. Finally, no sender type has an incentive to share a non-surprising signal if
$\pi_1-c_S\leq u_\emptyset$,
or equivalently if
$\pi_1\leq c_S+u_\emptyset$.
Hence, there exists a threshold $\bar{\pi}_1\in(0,1)$ such that the equilibrium
exists for any off-equilibrium belief $\pi_1\leq\bar{\pi}_1$.

So far we have characterized the equilibrium for $c_S < c_L$. We next consider $c_S\in[c_L,c_H]$. 
By Lemma \ref{lem:delta-mono}, $\Delta$ is strictly decreasing in
$\kappa(0,L,\tilde q)$, so $\Delta|_{\kappa=0}$ is its largest value. Since
$\Delta|_{\kappa=0} > 0$ if and only if $c_S < c_L$, for $c_S \geq c_L$ we have
$\Delta \leq 0$ throughout and the low ability sender remains silent,
$\kappa^*(0,L,\tilde q) = 0$. A shared surprising signal then originates only
from the high ability type sharing proper signals, so $\pi_{0P} = \pi_{0U} = 1$
and she shares if and only if $1 - c_S \geq \pi_\emptyset|_{\kappa=0}$. This
defines $c_H \equiv 1 - \pi_\emptyset|_{\kappa=0} \in (c_L, 1)$. Hence the high
ability sender shares every proper surprising signal for $c_S \in [c_L, c_H]$
and remains silent for $c_S > c_H$. \hfill\qed


\subsection{Proof of Fact \ref{fact:1}}
Without loss of generality consider $\sigma = 0$. The sender's belief that
$\sigma = 0$ is fake is $\tilde q^S = z^S_{0F}/z^S_0$. Since $z^S_0$ is decreasing in $p_S$ for all $\eta>\frac12$ while
$z^S_{0F}$ is independent of $p_S$, we have
$\partial\tilde q^S/\partial p_S>0$, so $\tilde q^S$
increases in $|p_S - \sigma| = p_S$. The case $\sigma = 1$ is analogous.

Since $\partial\Delta/\partial c_S < 0$ and $\partial\Delta/\partial p_S < 0$,
while $\partial\Delta/\partial\kappa < 0$ by Lemma \ref{lem:delta-mono},
the indifference condition $\Delta = 0$ yields
$\partial\kappa^*(0,L,\tilde q)/\partial c_S < 0$ and
$\partial\kappa^*(0,L,\tilde q)/\partial p_S < 0$.\hfill\qed

\subsection{Proof of Proposition \ref{prop:dgammadc}}
Recall that the quality of information is defined as $
\gamma=\sigma^\mathcal{F}/(\sigma^\mathcal{F}+\sigma^\mathcal{P})$. The probability that proper news is shared is $
\sigma^\mathcal{P}=(1-q)((1-p_T)\eta+p_T(1-\eta))\left[\lambda_S+(1-\lambda_S)\kappa_0\right]$,
whereas the probability that fake news is shared is $\sigma^\mathcal{F}=q (1-\lambda_S)(1-\beta)\kappa_0$. Moreover, from the proof of Fact \ref{fact:1} we know that
$\kappa_0^*$ decreases in $c_S$ and $p_S$, proving the proposition.\hfill\qed


\subsection{Proof of Proposition \ref{prop:dgammadbeta}}
Recall that 
$\gamma=\frac{\sigma^\mathcal{F}}{\sigma^\mathcal{F}+\sigma^\mathcal{P}}
$, 
with 
$\sigma^\mathcal{P}=(1-q)\left[  p_T(1-\eta )
 +(1-p_T)\eta  \right]\left((1-\lambda_S)\kappa_0^*+{\lambda_S}  \right)$ and $
 \sigma^\mathcal{F}= q(1-\beta)  (1-{\lambda_S}) \kappa_0^*$.
We know that $\gamma$ increases in $\kappa^*(0,L,\tilde{q})$. Hence, let $\kappa^*(0,L,\tilde{q})=1$. Then we have $
\left.\sigma^\mathcal{P}\right|_{\kappa^*(0,L,\tilde{q})=1}=(1-q)\left[  p_T(1-\eta )
 +(1-p_T)\eta  \right]$
and $\left.\sigma^\mathcal{F}\right|_{\kappa^*(0,L,\tilde{q})=1}=q(1-\beta)  (1-{\lambda_S})$, 
and hence $
\left.\gamma\right|_{\kappa^*(0,L,\tilde{q})=1}=\frac{q(1-\beta)  (1-{\lambda_S})}{(1-q)\left[  p_T(1-\eta )
 +(1-p_T)\eta  \right]+q(1-\beta)  (1-{\lambda_S})}$.
This is monotone decreasing in $\beta$
as $
\frac{\partial \left.\gamma\right|_{\kappa^*(0,L,\tilde{q})=1}}{\partial \beta}=\frac{(1-\lambda) (1-q) q ((2 \eta -1)
   p_T-\eta )}{(\eta +(2 \eta -1)
   p_T (q-1)-q (\beta  (-\lambda
   )+\beta +\eta +\lambda -1))^2}<0
   \Leftrightarrow (2 \eta -1)
   p_T-\eta <0$,
which is always true. To see this note that the LHS is maximized if $\eta=1$ when $p_T>\ot$ and when $\eta=\ot$ when $p_T<\ot$. In the first case, the LHS is $p_T-\eta<0$, in the latter $-p_T<0$. Finally, if $p_T=\ot$, the LHS equals $-\ot<0$.

To prove the proposition we hence only need to set $\left.\gamma\right|_{\kappa^*(0,L,\tilde{q})=1}=q$ and solve for $\beta$, which yields $
\tilde{\beta}=1-\frac{(1-p_T)\eta+p_T(1-\eta)}{1-\lambda_S}<1$. If $\beta>\tilde{\beta}$, then $\gamma<q$, independent of $\kappa^*(0,L,\tilde{q})$. This proves the proposition.
\qed

\subsection{Proof of Proposition \ref{Prop:Wview}}

In equilibrium it  has to hold that $C_l=C_h=0$ as defined in Conditions (\ref{wwIndif1}) and (\ref{wwIndif2}). 
We next prove that there always exists a constant $\xi\in(0,1)$ such that there exists $(p_{Sl},p_{Sh})\in(0,\xi)\times (\xi,1)$ that simultaneously solves  (\ref{wwIndif1}) and (\ref{wwIndif2}) if $c_S<\bar{c}_S\equiv\min\left\{1-\xi,\frac{\int_{0}^\xi z \,f(z)dz}{\int_{0}^\xi  \,f(z)dz}\right\}.$ If instead $c_S\geq\bar{c}_S$, we may have $p_{Sl}=0$ or $p_{Sh}=1$.


First consider $C_l$ and let $p_{Sl}=0$. Then:
\[
\left.C_l\right|_{p_{Sl}=0}=-c_S+\displaystyle\int_0^1\left|\frac{\widehat{p}_R \int_{0}^{p_{Sh}}z {f}_S(z)dz+(1-\widehat{p}_R) \int_{0}^1 z {f}_S(z)dz}{\widehat{p}_R {F}_S(p_{Sh})   +(1-\widehat{p}_R) (1-{F}_S(0))}\right|dF_R(p_R).
\]
This expression is strictly positive if $c_S=0$. We will now construct an upper bound on $c_S$ to ensure this expression is positive. Note that  $p_{Sh}\leq 1$, and therefore the expression in the integral weakly decreases in $\widehat{p}_R$. Thus, let $\widehat{p}_R=1$, yielding $
\left.C_l\right|_{p_{Sl}=0\wedge\widehat{p}_R=1}=-c_S+\frac{\int_{0}^{p_{Sh}}z {f}_S(z)dz}{{F}_S(p_{Sh})}$.
This expression increases in $p_{Sh}$, and hence if $p_{Sh}=\xi$, it is least likely to be positive. Letting $p_{Sh}=\xi$ gives us one of the conditions on $c_S$ from the proposition.

Next let $p_{Sl}=\xi$. Then:
\[
\begin{array}{rcl}
\left.C_l\right|_{p_{Sl}=\xi}&=&-\left(\xi-\dfrac{\int_0^\xi z f_S(z)dz}{\int_0^\xi  f_S(z)dz}\right)-c_S\\
&+&\displaystyle\int_0^1\left|\xi-\frac{\widehat{p}_R \int_{0}^{p_{Sh}}z {f}_S(z)dz+(1-\widehat{p}_R) \int_{\xi}^1 z {f}_S(z)dz}{\widehat{p}_R {F}_S(p_{Sh})   +(1-\widehat{p}_R) (1-{F}_S(\xi))}\right|dF_R(p_R).
\end{array}
\]
If the integral becomes very large, then this is not negative. 
How does $\hat{p}_S(\emptyset)$ change with $\hat{p}_R$?
\[
\dfrac{\partial \hat{p}_S(\emptyset)}{\partial \hat{p}_R}=\dfrac{\left(\int_{0}^{p_{Sh}}z {f}_S(z)dz\right)\left(1-F(p_{Sl})\right)-\left(\int_{p_{Sl}}^1 z {f}_S(z)dz\right)F(p_{Sh})}{\left(\widehat{p}_R {F}_S(p_{Sh})   +(1-\widehat{p}_R) (1-{F}_S(p_{Sl}))\right)^2}
\]
Thus, $\text{Sign}\left[\frac{\partial \hat{p}_S(\emptyset)}{\partial \hat{p}_R}\right]=\text{Sign}\left[\frac{\int_0^{p_{Sh}} z {f}_S(z)dz}{{F}_S(p_{Sh})}-\frac{\int_{p_{Sl}}^1 z {f}_S(z)dz}{(1-{F}_S(p_{Sl}))}\right]$.
Since $p_{Sh}\geq p_{Sl}$, $\hat{p}_S(\emptyset)$ is decreasing in $\hat{p}_R$, we can find an upper bound for the integral in $\left.C_l\right|_{p_{Sl}=\eta}$ either when $\hat{p}_R=1$ or when $\hat{p}_R=0$ for all $p_R$. 
If $\hat{p}_R=1$, then
\[
\begin{array}{rcl}
\left.C_l\right|_{p_{Sl}=\xi\wedge \hat{p}_R=1}&=&-\left(\xi-\dfrac{\int_0^\xi z f_S(z)dz}{\int_0^\xi  f_S(z)dz}\right)-c_S+\left|\xi-\dfrac{\int_{0}^{p_{Sh}}z {f}_S(z)dz}{{F}_S(p_{Sh})}\right|\\
&=&-\left(\xi-\dfrac{\int_0^\xi z f_S(z)dz}{\int_0^\xi  f_S(z)dz}\right)-c_S+\left(\xi-\dfrac{\int_{0}^{p_{Sh}}z {f}_S(z)dz}{{F}_S(p_{Sh})}\right)\\
&=&\dfrac{\int_0^\xi z f_S(z)dz}{\int_0^\xi  f_S(z)dz}-\dfrac{\int_{0}^{p_{Sh}}z {f}_S(z)dz}{{F}_S(p_{Sh})}-c_S\leq 0.
\end{array}
\]
The  inequality follows from 
$\dfrac{\int_0^{p_{Sh}} z f_S(z)dz}{\int_0^{p_{Sh}}  f_S(z)dz}\geq\dfrac{\int_0^\xi z f_S(z)dz}{\int_0^\xi  f_S(z)dz}$ because  $p_{Sh}\geq \xi$.
If instead $\hat{p}_R=0$, then
\[
\begin{array}{rcl}
\left.C_l\right|_{p_{Sl}=\xi\wedge \hat{p}_R=0}&=&-\left(\xi-\dfrac{\int_0^\xi z f_S(z)dz}{\int_0^\xi  f_S(z)dz}\right)+\left|\xi-\dfrac{\int_{\xi}^1 z {f}_S(z)dz}{(1-{F}_S(\xi))}\right|-c_S\\
&=&-\left(\xi-\dfrac{\int_0^\xi z f_S(z)dz}{\int_0^\xi  f_S(z)dz}\right)+\left(\dfrac{\int_{\xi}^1 z {f}_S(z)dz}{(1-{F}_S(\xi))}-\xi\right)-c_S.
\end{array}
\]
Hence, if 
\begin{equation}
\label{CC:CL_mu}
-2\xi+\dfrac{\int_0^\xi z f_S(z)dz}{F_S(\xi)}+\dfrac{\int_{\xi}^1 z {f}_S(z)dz}{(1-{F}_S(\xi))}-c_S\leq 0,    
\end{equation}
then there exists $\xi$ and $p_{Sl}$ such that $p_{Sl}\in(0,\xi)$ solves $C_l=0$ for all $p_{Sh}\in(\xi,1)$. 

Now  consider $C_h$ and let $p_{Sh}=1$:
\[
\left.C_h\right|_{p_{Sh}=1}= -c_S+\displaystyle\int_0^1\left|1-\frac{\widehat{p}_R \int_{0}^{p_{Sh}}z {f}_S(z)dz+(1-\widehat{p}_R) \int_{p_{Sl}}^1 z {f}_S(z)dz}{\widehat{p}_R {F}_S(p_{Sh})   +(1-\widehat{p}_R) (1-{F}_S(p_{Sl}))}\right|dF_R(p_R).
\]
If $c_S$ is small, this is positive. As before, the integral is minimized if either $\widehat{p}_R=0$ or  $\widehat{p}_R=1$. If  $\widehat{p}_R=0$, then  $\left.C_h\right|_{p_{Sh}=1\wedge \widehat{p}_R=0}= -c_S+1-\frac{\int_{p_{Sl}}^1 z {f}_S(z)dz}{(1-{F}_S(p_{Sl}))}$. This is smallest if $p_{Sl}=0$, in which case we get $
\left.C_h\right|_{p_{Sh}=1\wedge \widehat{p}_R=0 \wedge p_{Sl}=0}= -c_S+1-\xi$.
This is positive if $c_S<1-\xi$, which is the second condition stated in the proposition.

Finally,  let $p_{Sh}=\xi$:
\[
\left.C_h\right|_{p_{Sh}=\xi}=\xi-\dfrac{\int_\xi^1 z f_S(z)dz}{\int_\xi^1  f_S(z)dz}-c_S+\displaystyle\int_0^1\left|\xi-\frac{\widehat{p}_R \int_{0}^{\xi}z {f}_S(z)dz+(1-\widehat{p}_R) \int_{p_{Sl}}^1 z {f}_S(z)dz}{\widehat{p}_R {F}_S(\xi)   +(1-\widehat{p}_R) (1-{F}_S(p_{Sl}))}\right|dF_R(p_R)
\]
The expression is maximized if either $\widehat{p}_R=0$ or  $\widehat{p}_R=1$. 
Letting $\widehat{p}_R=0$,
\[
\begin{array}{rcl}
\left.C_h\right|_{p_{Sh}=\xi\wedge \hat{p}_R=0}&=&-\left(\dfrac{\int_\xi^1 z f_S(z)dz}{\int_\xi^1  f_S(z)dz}-\xi\right)-c_S+\left|\xi-\dfrac{\int_{p_{Sl}}^1 z {f}_S(z)dz}{(1-{F}_S(p_{Sl}))}\right| \\
&=&-\left(\dfrac{\int_\xi^1 z f_S(z)dz}{\int_\xi^1  f_S(z)dz}-\xi\right)-c_S+\left(\dfrac{\int_{p_{Sl}}^1 z {f}_S(z)dz}{(1-{F}_S(p_{Sl}))}-\xi\right)\\
&=&\dfrac{\int_{p_{Sl}}^1 z {f}_S(z)dz}{(1-{F}_S(p_{Sl}))}-\dfrac{\int_{\xi}^1 z {f}_S(z)dz}{(1-{F}_S(\xi))}-c_S<0
\end{array}
\]
If instead  $\hat{p}_R=1$, then
\[
\begin{array}{rcl}
\left.C_h\right|_{p_{Sh}=\xi\wedge \hat{p}_R=1}&=&-\left(\dfrac{\int_\xi^1 z f_S(z)dz}{\int_\xi^1  f_S(z)dz}-\xi\right)-c_S+\left|\xi-\dfrac{\int_{0}^{\xi}z {f}_S(z)dz}{{F}_S(\xi)   }\right| \\
&=&-\left(\dfrac{\int_\xi^1 z f_S(z)dz}{\int_\xi^1  f_S(z)dz}-\xi\right)-c_S+\left(\xi-\dfrac{\int_{0}^{\xi}z {f}_S(z)dz}{{F}_S(\xi)   }\right).
\end{array}
\]
Hence, if 
\begin{equation}
\label{CC:CH_mu}
2\xi -\dfrac{\int_\xi^1 z f_S(z)dz}{\int_\xi^1  f_S(z)dz}-\dfrac{\int_{0}^{\xi}z {f}_S(z)dz}{{F}_S(\xi)   }-c_S\leq0
\end{equation}
and \eqref{CC:CL_mu},  
then there exist $\xi$ and $p_{Sh}$ such that $p_{Sh}\in(\xi,1)$ solves $C_h=0$ for any $p_{Sl}\in(0,\xi)$. Therefore, what is left to prove is  that there exists  $\xi\in(0,1)$ guaranteeing that both \eqref{CC:CL_mu} and \eqref{CC:CH_mu} are satisfied simultaneously. This is indeed the case for all $c_S\geq 0$ if $
2\xi -\frac{\int_\xi^1 z f_S(z)dz}{\int_\xi^1  f_S(z)dz}-\frac{\int_{0}^{\xi}z {f}_S(z)dz}{{F}_S(\xi)   }=0$. 
If $\xi=0$, then the LHS is $-\mathbb{E}[p_S]<0$, while if $\xi=1$ the LHS is $2-\mathbb{E}[p_S]>0$. It follows from continuity of the LHS that there exists $\xi\in(0,1)$ such that the equation is satisfied. Because $[0,\xi]\times[\xi,1]$ is compact, and because both $C_l$ and $C_h$ are continuous, it follows that there must exist
$(p_{Sl},p_{Sh})\in[0,\xi]\times[\xi,1]$ such that $C_l=C_h=0$. 
\qed


\subsection{Proof of Proposition \ref{Prop:Receiver_WW_pR}}
\label{app:proofReceiver_WW_pR}
First, we prove a lemma that is important to simplify the analysis:
\begin{lem}
\label{lemma:ranking}
In any interior equilibrium with known receiver type, it holds that
\begin{equation}
\begin{array}{c}
\left|\frac{\widehat{p}_R \int_{0}^{p_{Sh}^*}z {f}_S(z)dz+(1-\widehat{p}_R) \int_{p_{Sl}^*}^1 z {f}_S(z)dz}{\widehat{p}_R {F}_S(p_{Sh}^*)   +(1-\widehat{p}_R) (1-{F}_S(p_{Sl}^*))}-p_{Sl}^*\right|
=\frac{\widehat{p}_R \int_{0}^{p_{Sh}^*}z {f}_S(z)dz+(1-\widehat{p}_R) \int_{p_{Sl}^*}^1 z {f}_S(z)dz}{\widehat{p}_R {F}_S(p_{Sh}^*)   +(1-\widehat{p}_R) (1-{F}_S(p_{Sl}^*))}-p_{Sl}^*
\label{eq:lem:line1}    
\end{array}
\end{equation}
and 
\begin{equation}
\begin{array}{c}
\left|p_{Sh}^*-\frac{\widehat{p}_R \int_{0}^{p_{Sh}^*}z {f}_S(z)dz+(1-\widehat{p}_R) \int_{p_{Sl}^*}^1 z {f}_S(z)dz}{\widehat{p}_R {F}_S(p_{Sh}^*)   +(1-\widehat{p}_R) (1-{F}_S(p_{Sl}^*))}\right|=p_{Sh}^*-\frac{\widehat{p}_R \int_{0}^{p_{Sh}^*}z {f}_S(z)dz+(1-\widehat{p}_R) \int_{p_{Sl}^*}^1 z {f}_S(z)dz}{\widehat{p}_R {F}_S(p_{Sh}^*)   +(1-\widehat{p}_R) (1-{F}_S(p_{Sl}^*))}.
\end{array}
\label{eq:lem:line2}    
\end{equation}

\end{lem}
\begin{proof}
To prove \eqref{eq:lem:line1}, recall that $\widehat{p}_S(\emptyset)$ is monotone in $\hat{p}_R$. If $\hat{p}_R=0$, the LHS of \eqref{eq:lem:line1} becomes
$
\left|\frac{\int_{p_{Sl}^*}^1 z f_S(z)dz}{1-F_S(p_{Sl}^*)}-p_{Sl}^*\right|=\frac{\int_{p_{Sl}^*}^1 z f_S(z)dz}{1-F_S(p_{Sl}^*)}-p_{Sl}^*>0$. 
If $\hat{p}_R=1$, it becomes $
\left|\frac{\int_0^{p_{Sh}^*} z f_S(z)dz}{F_S(p_{Sh}^*)}-p_{Sl}^*\right|$.
If the interior of the absolute value is also positive, we proved \eqref{eq:lem:line1}. Thus, assume it was negative. Then $C_l$ becomes $\frac{\int_0^{p_{Sl}^*} z f_S(z)dz}{F_S(p_{Sl}^*)}-\frac{\int_0^{p_{Sh}^*} z f_S(z)dz}{F_S(p_{Sh}^*)}-c_S<0$. This is negative, and hence the equilibrium condition can never be satisfied with known receiver worldview. Thus, for known $p_R$, the interior of the absolute value must be positive in equilibrium. The proof of \eqref{eq:lem:line2} follows the same line of reasoning.
\end{proof}

From the lemma it follows that in a  Responsive Equilibrium with one receiver with known type, the equilibrium conditions in (\ref{wwIndif1}) and (\ref{wwIndif2}) simplify to
\begin{eqnarray}
\begin{array}{rcl}
C_l&=&\hat{p}_S(0)+ \hat{p}_S(\emptyset)- 2p_{Sl}-c_S,\\
C_h&=&2p_{Sh} -\hat{p}_S(1)-\hat{p}_S(\emptyset)-c_S.
\end{array}    
\end{eqnarray}
We know from the proof of Proposition \ref{Prop:Wview} that $C_l$ is positive for $p_{Sl}=0$ and negative for $p_{Sl}=\xi$. From Assumption \ref{assume:Jacobian} it follows that  $\frac{\partial C_l}{\partial p_{Sl}}=\frac{\partial \hat{p}_S(0)}{\partial p_{Sl}}+\frac{\partial  \hat{p}_S(\emptyset)}{\partial p_{Sl}}- 2<0$ for all $p_{Sl}$, see Assumption \ref{assume:Jacobian}. Similarly, we know from the proof of Proposition \ref{Prop:Wview} that $C_h$ is negative for $p_{Sh}=\xi$ and positive for $p_{Sh}=1$. Thus, to guarantee  uniqueness of equilibrium, we assume $\frac{\partial C_h}{\partial p_{Sh}}=2-\frac{\partial \hat{p}_S(1)}{\partial p_{Sh}}+\frac{\partial  \hat{p}_S(\emptyset)}{\partial p_{Sh}}>0$ for all $p_{Sh}$ (Assumption \ref{assume:Jacobian} as well). Thus,we assume that beliefs do not change too abruptly with $p_{Sl}$ and $p_{Sh}$. 

Under Assumption \ref{assume:Jacobian} there is a 
unique $(p_{Sl},p_{Sh})\in[0,\xi]\times [\xi,1]$ solving the equilibrium conditions (
\citealp{GaleNikaido:1965}). 
To derive comparative statics,  define
\[
\begin{array}{ccc}
 J=
  \left(
  \begin{array}{cc}
  \frac{\partial C_l}{\partial p_{Sl}}&\frac{\partial C_l}{\partial p_{Sh}}\\
  \frac{\partial C_h}{\partial p_{Sl}}&\frac{\partial C_h}{\partial p_{Sh}}
  \end{array}\right)  &
 J_L^{p_R}=
  \left(
  \begin{array}{cc}
  -\frac{\partial C_l}{\partial p_R}&\frac{\partial C_l}{\partial p_{Sh}}\\
  -\frac{\partial C_h}{\partial p_R}&\frac{\partial C_h}{\partial p_{Sh}}
  \end{array}\right)  &
  J_H^{p_R}=
  \left(
  \begin{array}{cc}
  \frac{\partial C_l}{\partial p_{Sl}}&-\frac{\partial C_l}{\partial p_R}\\
  \frac{\partial C_h}{\partial p_{Sl}}&-\frac{\partial C_h}{\partial p_R}
  \end{array}\right)  \\
  &\\
 J_L^c=
  \left(
  \begin{array}{cc}
  -\frac{\partial C_l}{\partial c_S}&\frac{\partial C_l}{\partial p_{Sh}}\\
  -\frac{\partial C_h}{\partial c_S}&\frac{\partial C_h}{\partial p_{Sh}}
  \end{array}\right)  &
  J_H^c=
  \left(
  \begin{array}{cc}
  \frac{\partial C_l}{\partial p_{Sl}}&-\frac{\partial C_l}{\partial c_S}\\
  \frac{\partial C_h}{\partial p_{Sl}}&-\frac{\partial C_h}{\partial c_S}
  \end{array}\right).  
\end{array}
\]
Then
\[
\begin{array}{rcl}
\dfrac{\partial p_{Sl}}{\partial p_R}&=&\dfrac{\left|J_L^{p_R}\right|}{\left|J\right|}=\frac{-\frac{\partial C_l}{\partial p_R}\frac{\partial C_h}{\partial p_{Sh}}+\frac{\partial C_l}{\partial p_{Sh}}\frac{\partial C_h}{\partial p_R}}{\left|J\right|}=\frac{-\frac{\partial \hat{p}_S(\emptyset)}{\partial p_R}\left(\frac{\partial C_h}{\partial p_{Sh}}+\frac{\partial C_l}{\partial p_{Sh}}\right)}{\left|J\right|}\\
\dfrac{\partial p_{Sh}}{\partial p_R}&=&\dfrac{\left|J_H^{p_R}\right|}{\left|J\right|}=\frac{-\frac{\partial C_h}{\partial p_R}\frac{\partial C_l}{\partial p_{Sl}}+\frac{\partial C_h}{\partial p_{Sl}}\frac{\partial C_l}{\partial p_R}}{\left|J\right|}=\frac{\frac{\partial \hat{p}_S(\emptyset)}{\partial p_R}\left(\frac{\partial C_l}{\partial p_{Sl}}+\frac{\partial C_h}{\partial p_{Sl}}\right)}{\left|J\right|}\\
\dfrac{\partial p_{Sl}}{\partial c_S}&=&\dfrac{\left|J_L^c\right|}{\left|J\right|}=\frac{-\frac{\partial C_l}{\partial c_S}\frac{\partial C_h}{\partial p_{Sh}}+\frac{\partial C_l}{\partial p_{Sh}}\frac{\partial C_h}{\partial c_S}}{\left|J\right|}=\frac{\frac{\partial C_h}{\partial p_{Sh}}-\frac{\partial C_l}{\partial p_{Sh}}}{\left|J\right|}\\
\dfrac{\partial p_{Sh}}{\partial c_S}&=&\dfrac{\left|J_H^c\right|}{\left|J\right|}=\frac{-\frac{\partial C_h}{\partial c_S}\frac{\partial C_l}{\partial p_{Sl}}+\frac{\partial C_h}{\partial p_{Sl}}\frac{\partial C_l}{\partial c_S}}{\left|J\right|}=\frac{\frac{\partial C_l}{\partial p_{Sl}}-\frac{\partial C_h}{\partial p_{Sl}}}{\left|J\right|},
\end{array}
\]
where we used $\frac{\partial C_l}{\partial p_R}=-\frac{\partial C_h}{\partial p_R}=\frac{\partial\hat{p}_S(\emptyset)}{\partial p_R}$ and $\frac{\partial C_l}{\partial c_S}=\frac{\partial C_h}{\partial c_S}=-1$. In the proof of Proposition \ref{Prop:Wview} we have shown that $\frac{\partial\hat{p}_S(\emptyset)}{\partial p_R}<0$ and Assumption \ref{assume:Jacobian} states that $|J|<0$. This implies that 
$\text{Sign}\left[\frac{\partial p_{Sl}}{\partial p_R}\right]=\text{Sign}\left[-\left(\frac{\partial C_h}{\partial p_{Sh}}+\frac{\partial C_l}{\partial p_{Sh}}\right)\right]$, $
\text{Sign}\left[\frac{\partial p_{Sh}}{\partial p_R}\right]=\text{Sign}\left[\left(\frac{\partial C_l}{\partial p_{Sl}}+\frac{\partial C_h}{\partial p_{Sl}}\right)\right]$, $\text{Sign}\left[\frac{\partial p_{Sl}}{\partial c_S}\right]=\text{Sign}\left[\frac{\partial C_l}{\partial p_{Sh}}-\frac{\partial C_h}{\partial p_{Sh}}\right]$, and  
$\text{Sign}\left[\frac{\partial p_{Sh}}{\partial c_S}\right]=\text{Sign}\left[\frac{\partial C_h}{\partial p_{Sl}}-\frac{\partial C_l}{\partial p_{Sl}}\right]$. 
Next note that $\frac{\partial C_l}{\partial p_{Sh}}=\frac{\partial \hat{p}_S(\emptyset)}{\partial p_{Sh}}>0$ and $\frac{\partial C_h}{\partial p_{Sl}}=-\frac{\partial \hat{p}_S(\emptyset)}{\partial p_{Sl}}<0$. Together with our assumptions of $\frac{\partial C_l}{\partial p_{Sl}}<0$ and $\frac{\partial C_h}{\partial p_{Sh}}>0$, we can sign $\frac{\partial p_{Sl}}{\partial p_R}<0$ and $\frac{\partial p_{Sh}}{\partial p_R}<0$. Thus, both equilibrium thresholds decrease in $p_R$.
Moreover, $|J|<0$, $\frac{\partial C_l}{\partial p_{Sl}}<0$ and $\frac{\partial C_h}{\partial p_{Sh}}>0$ imply that $\frac{\partial p_{Sh}}{\partial c_S}>0$ and $\frac{\partial p_{Sl}}{\partial c_S}<0$. Thus, higher costs of sharing decrease the amount of signals shared from both types of signals.

\subsection{Proof of Proposition \ref{Prop:WorldviewFakeGeneral}}

Let $\beta=\hat{\beta}\equiv p_T \eta+(1-p_T)(1-\eta)\in(0,1)$. 
Then \eqref{Eq:ShareFake} reduces to $\gamma=q$. Next, notice that the only expressions in  \eqref{Eq:ShareFake} that depend on $p_R$ are  $F_S(p_{Sl})$ and $F_S(p_{Sh})$, implicitly  through the thresholds $p_{Sl}$ and $p_{Sh}$, respectively. 
Hence, the derivative of \eqref{Eq:ShareFake} is
\[
\begin{array}{rcl}
\dfrac{\partial \gamma}{\partial p_R}&=&-\dfrac{(1-q) q (\beta -\hat{\beta})
   }{\left({F_S(p_{Sl})\mathbb{P}(\sigma=0)+(1-F_S(p_{Sh}))\mathbb{P}(\sigma=1)}\right)^2} \\ 
   &&\times \left[F_S(p_{Sl}) {f_S}(p_{Sh})
   \frac{\partial p_{Sh}}{\partial p_R}+{f_S}(p_{Sl})
   (1-F_S(p_{Sh}))
    \frac{\partial p_{Sl}}{\partial p_R}\right].
  \end{array}
\]
Clearly, if $\beta=\hat{\beta}$, this derivative equals zero. If $\beta>\hat{\beta}$, the derivative is positive iff the expression in squared parentheses is negative. We know from Proposition \ref{Prop:Receiver_WW_pR} that both $p_{Sl}$ and $p_{Sh}$ are decreasing in $p_R$, and hence the derivative is indeed positive when $\beta>\hat{\beta}$. Similarly, if $\beta<\hat{\beta}$, the derivative is negative. This proves items (i) and (ii) of the proposition.\qed
%

\subsection{Proof of Proposition \ref{Prop:WvInfDet}}
That $q=\gamma$ when $\beta=\hat{\beta}$ follows from Proposition \ref{Prop:WorldviewFakeGeneral}.
For the other cases, recall that \begin{eqnarray*}
    \gamma-q=q\left(\frac{(1-\beta)F_S(p_{Sl})+\beta(1-F_S(p_{Sh}))}{\mathbb{P}(\sigma=0)F_S(p_{Sl})+\mathbb{P}(\sigma=1)(1-F_S(p_{Sh}))}-1\right),
\end{eqnarray*}
\normalsize
\noindent
where $\mathbb{P}(\sigma=1)=q\beta+(1-q)\left[p_T \eta+(1-p_T)(1-\eta)\right]$
and $\mathbb{P}(\sigma=0)=1-\mathbb{P}(\sigma=1)$. Thus, $\gamma>q$ iff
\[
\begin{array}{c}
    (1-\beta)F_S(p_{Sl})+\beta(1-F_S(p_{Sh}))-\mathbb{P}(\sigma=0)F_S(p_{Sl})-\mathbb{P}(\sigma=1)(1-F_S(p_{Sh}))>0\\
  \Leftrightarrow  (\hat{\beta}-\beta)\left(F_S(p_{Sl})-(1-F_S(p_{Sh}))\right)>0.
\end{array}
\]
This proves the proposition.\qed


\begin{thebibliography}{}

\bibitem[Acemoglu et~al., 2010]{AcemogluEtAl:2010}
Acemoglu, D., Ozdaglar, A., and Gheibi, A.~P. (2010).
\newblock Spread of (mis)information in social networks.
\newblock {\em Games and Economic Behavior}, 70(2):194 -- 227.

\bibitem[Acemoglu et~al., 2023]{AcemogluOzdaglarSiderius:2023}
Acemoglu, D., Ozdaglar, A., and Siderius, J. (2023).
\newblock {A model of online misinformation}.
\newblock {\em The Review of Economic Studies}, 91:3117--3150.

\bibitem[Allcott and Gentzkow, 2017]{allcott2017social}
Allcott, H. and Gentzkow, M. (2017).
\newblock Social media and fake news in the 2016 election.
\newblock {\em Journal of Economic Perspectives}, 31(2):211--36.

\bibitem[Altman and Taylor, 1973]{altman1973social}
Altman, I. and Taylor, D.~A. (1973).
\newblock {\em Social penetration: The development of interpersonal
  relationships.}
\newblock Holt, Rinehart \& Winston.

\bibitem[Angelucci and Prat, 2024]{AngelucciPratt:2021}
Angelucci, C. and Prat, A. (2024).
\newblock {Is journalistic truth dead? Measuring how informed voters are about
  political news}.
\newblock {\em American Economic Review}, 114(4):887--925.

\bibitem[Banerjee et~al., 2013]{BanerjeeEtAl:2013}
Banerjee, A., Chandrasekhar, A.~G., Duflo, E., and Jackson, M.~O. (2013).
\newblock The diffusion of microfinance.
\newblock {\em Science}, 341(6144):1236498.

\bibitem[Banerjee et~al., 2019]{BanerjeeEtAl:2019}
Banerjee, A., Chandrasekhar, A.~G., Duflo, E., and Jackson, M.~O. (2019).
\newblock {Using gossips to spread information: Theory and evidence from two
  randomized controlled trials}.
\newblock {\em The Review of Economic Studies}, 86(6):2453--2490.

\bibitem[Bazarova and Choi, 2014]{bazarova2014self}
Bazarova, N.~N. and Choi, Y.~H. (2014).
\newblock Self-disclosure in social media: Extending the functional approach to
  disclosure motivations and characteristics on social network sites.
\newblock {\em Journal of Communication}, 64(4):635--657.

\bibitem[Bowen et~al., 2023]{BowenEtAl:2023}
Bowen, T.~R., Dmitriev, D., and Galperti, S. (2023).
\newblock {Learning from shared news: When abundant information leads to belief
  polarization}.
\newblock {\em The Quarterly Journal of Economics}, 138(2):955--1000.

\bibitem[Buisseret and Van~Weelden, 2022]{buisseret2022polarization}
Buisseret, P. and Van~Weelden, R. (2022).
\newblock Polarization, valence, and policy competition.
\newblock {\em American Economic Review: Insights}, 4(3):341--352.

\bibitem[Cantarella et~al., 2023]{cantarella2023does}
Cantarella, M., Fraccaroli, N., and Volpe, R. (2023).
\newblock Does fake news affect voting behaviour?
\newblock {\em Research Policy}, 52(1):104628.

\bibitem[Che and Kartik, 2009]{CheKartik:2009}
Che, Y. and Kartik, N. (2009).
\newblock Opinions as incentives.
\newblock {\em Journal of Political Economy}, 117(5):815--860.

\bibitem[Crawford and Sobel, 1982]{CrawfordSobel:1982}
Crawford, V.~P. and Sobel, J. (1982).
\newblock Strategic information transmission.
\newblock {\em Econometrica}, 50(6):1431--1451.

\bibitem[Denter et~al., 2021]{DDG:2021}
Denter, P., Dumav, M., and Ginzburg, B. (2021).
\newblock Social connectivity, media bias, and correlation neglect.
\newblock {\em The Economic Journal}, 131:2033--2057.

\bibitem[Denter and Ginzburg, 2026]{DenterGinzburg:2026}
Denter, P. and Ginzburg, B. (2026).
\newblock Troll farms.
\newblock Technical report, arXiv:2411.03241.

\bibitem[Downs, 1957]{downs1957economic}
Downs, A. (1957).
\newblock An economic theory of political action in a democracy.
\newblock {\em Journal of political economy}, 65(2):135--150.

\bibitem[Enkhbayar et~al., 2020]{enkhbayar2020much}
Enkhbayar, A., Haustein, S., Barata, G., and Alperin, J.~P. (2020).
\newblock How much research shared on {F}acebook happens outside of public
  pages and groups? {A} comparison of public and private online activity around
  {PLOS ONE} papers.
\newblock {\em Quantitative Science Studies}, 1(2):749--770.

\bibitem[Erdt et~al., 2016]{erdt2016altmetrics}
Erdt, M., Nagarajan, A., Sin, S.-C.~J., and Theng, Y.-L. (2016).
\newblock Altmetrics: an analysis of the state-of-the-art in measuring research
  impact on social media.
\newblock {\em Scientometrics}, 109(2):1117--1166.

\bibitem[Gale and Nikaido, 1965]{GaleNikaido:1965}
Gale, D. and Nikaido, H. (1965).
\newblock The jacobian matrix and global univalence of mappings.
\newblock {\em Mathematische Annalen}, 159:81--93.

\bibitem[Germano et~al., 2026]{germano2026ranking}
Germano, F., G{\'o}mez, V., and Sobbrio, F. (2026).
\newblock Ranking for engagement: How social media algorithms fuel
  misinformation and polarization.
\newblock {\em Journal of Public Economics}, 255:105589.

\bibitem[Greene et~al., 2006]{greene2006self}
Greene, K., Derlega, V.~J., and Mathews, A. (2006).
\newblock Self-disclosure in personal relationships.
\newblock {\em The Cambridge handbook of personal relationships}, 409:427.

\bibitem[Grossman and Helpman, 2023]{GrossmanHelpman:2019}
Grossman, G.~M. and Helpman, E. (2023).
\newblock Electoral competition with fake news.
\newblock {\em European Journal of Political Economy}, 77:102315.

\bibitem[Guess et~al., 2019]{GuessEtAl:2019}
Guess, A., Nagler, J., and Tucker, J. (2019).
\newblock {Less than you think: Prevalence and predictors of fake news
  dissemination on Facebook}.
\newblock {\em Science Advances}, 5(1).

\bibitem[Guess et~al., 2023]{GuessEtAl:2023}
Guess, A.~M., Malhotra, N., Pan, J., Barber\'a, P., Allcott, H., Brown, T.,
  Crespo-Tenorio, A., Dimmery, D., Freelon, D., Gentzkow, M.,
  Gonz\'alez-Bail\'on, S., Kennedy, E., Kim, Y.~M., Lazer, D., Moehler, D.,
  Nyhan, B., Rivera, C.~V., Settle, J., Thomas, D.~R., Thorson, E., Tromble,
  R., Wilkins, A., Wojcieszak, M., Xiong, B., de~Jonge, C.~K., Franco, A.,
  Mason, W., Stroud, N.~J., and Tucker, J.~A. (2023).
\newblock Reshares on social media amplify political news but do not detectably
  affect beliefs or opinions.
\newblock {\em Science}, 381(6656):404--408.

\bibitem[Guriev et~al., 2025]{GurievHenryMarquisZhuravskaya:2023}
Guriev, S., Henry, E., Marquis, T., and Zhuravskaya, E. (2025).
\newblock Curtailing false news, amplifying truth.
\newblock {CEPR Discussion Paper No. 18650}, CEPR.

\bibitem[Kranton and McAdams, 2024]{KrantonMcAdams:2024}
Kranton, R. and McAdams, D. (2024).
\newblock Social connectedness and information markets.
\newblock {\em American Economic Journal: Microeconomics}, 16:33--62.

\bibitem[K\"{u}mpel et~al., 2015]{KuempelEtAl:2015}
K\"{u}mpel, A.~S., Karnowski, V., and Keyling, T. (2015).
\newblock News sharing in social media: A review of current research on news
  sharing users, content, and networks.
\newblock {\em Social Media + Society}, 1(2):1--14.

\bibitem[Lee and Ma, 2012]{LeeMa:2012}
Lee, C.~S. and Ma, L. (2012).
\newblock News sharing in social media: The effect of gratifications and prior
  experience.
\newblock {\em Computers in Human Behavior}, 28(2):331--339.

\bibitem[Lee et~al., 2011]{LeeMaGoh:2011}
Lee, C.~S., Ma, L., and Goh, D. H.-L. (2011).
\newblock Why do people share news in social media?
\newblock In Zhong, N., Callaghan, V., Ghorbani, A.~A., and Hu, B., editors,
  {\em Active Media Technology}, pages 129--140, Berlin, Heidelberg. Springer
  Berlin Heidelberg.

\bibitem[Levy and Razin, 2018]{LEVY2018262}
Levy, G. and Razin, R. (2018).
\newblock Information diffusion in networks with the bayesian peer influence
  heuristic.
\newblock {\em Games and Economic Behavior}, 109:262--270.

\bibitem[Lin et~al., 2023]{lin2023thinking}
Lin, H., Pennycook, G., and Rand, D.~G. (2023).
\newblock Thinking more or thinking differently? using drift-diffusion modeling
  to illuminate why accuracy prompts decrease misinformation sharing.
\newblock {\em Cognition}, 230:105312.

\bibitem[Marwick and boyd, 2011]{marwick2011tweet}
Marwick, A.~E. and boyd, d. (2011).
\newblock I tweet honestly, i tweet passionately: Twitter users, context
  collapse, and the imagined audience.
\newblock {\em New Media \& Society}, 13(1):114--133.

\bibitem[Montagni et~al., 2021]{MontagniEtAl:2021}
Montagni, I., Ouazzani-Touhami, K., Mebarki, A., Texier, N., Sch{\"u}ck, S.,
  Tzourio, C., and the CONFINS~group (2021).
\newblock {Acceptance of a Covid-19 vaccine is associated with ability to
  detect fake news and health literacy}.
\newblock {\em Journal of Public Health}, 43(4):695--702.

\bibitem[Osmundsen et~al., 2021]{OsmundsenEtAl:2021}
Osmundsen, M., Bor, A., Vahlstrup, P.~B., Bechmann, A., and Petersen, M.~B.
  (2021).
\newblock Partisan polarization is the primary psychological motivation behind
  political fake news sharing on twitter.
\newblock {\em American Political Science Review}, 115(3):999--1015.

\bibitem[Papanastasiou, 2020]{Papanastasiou:2020}
Papanastasiou, Y. (2020).
\newblock Fake news propagation and detection: A sequential model.
\newblock {\em Management Science}, 66(5):1826--1846.

\bibitem[Park et~al., 2020]{park2020trust}
Park, K., Kwak, H., Song, H., and Cha, M. (2020).
\newblock ``trust me, i have a ph. d.'': A propensity score analysis on the
  halo effect of disclosing one's offline social status in online communities.
\newblock In {\em Proceedings of the international AAAI conference on web and
  social media}, volume~14, pages 534--544.

\bibitem[Pennycook et~al., 2021]{PennycookEtAl:2021}
Pennycook, G., Epstein, Z., Mosleh, M., Arechar, A.~A., Eckles, D., and Rand,
  D.~G. (2021).
\newblock Shifting attention to accuracy can reduce misinformation online.
\newblock {\em Nature}, 592:590 -- 595.

\bibitem[Pogorelskiy and Shum, 2019]{pogorelskiy2019news}
Pogorelskiy, K. and Shum, M. (2019).
\newblock News we like to share: How news sharing on social networks influences
  voting outcomes.
\newblock Technical report.

\bibitem[Stein et~al., 2024]{stein2024partisan}
Stein, J., Keuschnigg, M., and van~de Rijt, A. (2024).
\newblock Partisan belief in new misinformation is resistant to accuracy
  incentives.
\newblock {\em PNAS nexus}, 3(11):pgae506.

\bibitem[Sundar et~al., 2025]{SundarEtAl:2025}
Sundar, S.~S., Snyder, E.~C., Liao, M., Yin, J., Wang, J., and Chi, G. (2025).
\newblock Sharing without clicking on news in social media.
\newblock {\em Nature Human Behavior}, 9:156--168.

\bibitem[Taber and Lodge, 2006]{TaberLodge:2006}
Taber, C.~S. and Lodge, M. (2006).
\newblock Motivated skepticism in the evaluation of political beliefs.
\newblock {\em American Journal of Political Science}, 50(3):755--769.

\bibitem[van~der Does et~al., 2022]{vanDerDoes2022strategic}
van~der Does, T., Galesic, M., Dunivin, Z.~O., and Smaldino, P.~E. (2022).
\newblock Strategic identity signaling in heterogeneous networks.
\newblock {\em Proceedings of the National Academy of Sciences},
  119(10):e2117898119.

\bibitem[Vignery and Laurier, 2020]{VIGNERY2020101499}
Vignery, K. and Laurier, W. (2020).
\newblock Achievement in student peer networks: A study of the selection
  process, peer effects and student centrality.
\newblock {\em International Journal of Educational Research}, 99:101499.

\bibitem[Vosoughi et~al., 2018]{VosoughiRoyAral:2018}
Vosoughi, S., Roy, D., and Aral, S. (2018).
\newblock The spread of true and false news online.
\newblock {\em Science}, 359(6380):1146--1151.

\bibitem[Williams et~al., 2019]{williams2019linking}
Williams, E.~A., Zwolak, J.~P., Dou, R., and Brewe, E. (2019).
\newblock Linking engagement and performance: The social network analysis
  perspective.
\newblock {\em Physical review physics education research}, 15(2):020150.

\bibitem[Zimmermann and Kohring, 2020]{zimmermann2020mistrust}
Zimmermann, F. and Kohring, M. (2020).
\newblock Mistrust, disinforming news, and vote choice: A panel survey on the
  origins and consequences of believing disinformation in the 2017 german
  parliamentary election.
\newblock {\em Political communication}, 37(2):215--237.

\end{thebibliography}
\end{document}